\global\def\draftcontrol{0}

   \def\versionno{Universality of Annulons }

\catcode`\@=11

\expandafter\ifx\csname draftcontrol\endcsname\relax\global\def\draftcontrol{0}
\fi

{\count255=\time\divide\count255 by 60
\xdef\hourmin{\number\count255}
\multiply\count255 by-60\advance\count255 by\time
\xdef\hourmin{\hourmin:\ifnum\count255<10 0\fi\the\count255}}
\def\draftdate{\number\month/\number\day/\number\year\ \ \ \hourmin }

\newcommand\makepapertitle{\par
  \begingroup \renewcommand\thefootnote{\@fnsymbol\c@footnote}%
    \def\@makefnmark{\rlap{\@textsuperscript{\normalfont\@thefnmark}}}%
    \long\def\@makefntext##1{\parindent 1em\noindent \hb@xt@1.8em{%
    \hss\@textsuperscript{\normalfont\@thefnmark}}##1}
    \global\@topnum\z@   
    \@makepapertitle \thispagestyle{empty}\@thanks \endgroup
    \setcounter{footnote}{0}
    \global\let\makepapertitle\relax \global\let\@makepapertitle\relax
    \global\let\@thanks\@empty \global\let\@author\@empty
    \global\let\@date\@empty \global\let\@title\@empty
    \global\let\title\relax \global\let\author\relax
    \global\let\date\relax \global\let\and\relax
    \def\version{\let\version\@version\@gobble} }
\def\@makepapertitle{%
  \newpage \ifnum\draftcontrol=1 {} \version\versionno \vskip 3em%
   \else \hfill\hbox to 3cm {\parbox{4cm}{\@pubnum}\hss}
   \fi \begin{center}
   \vskip 1.5em
   \begin{tabular}[t]{c}
   {\@bstract}
}

\gdef\@pubnum{}
\def\pubnum#1{%
  \gdef\@pubnum{#1}}

\gdef\@bstract{}
\def\Abstract#1{%
  \gdef\@bstract{%
   \parbox{\textwidth-0pc}{%
   \centerline{\bf Abstract}\penalty1000%
\noindent
\renewcommand\baselinestretch{1.0}%
{#1}}}}

\def\ps@paper{\let\@mkboth\@gobbletwo%
     \ifnum\draftcontrol=1
        \def\@oddfoot{\hbox to \textwidth{\tiny \versionno \hfil\tiny\draftdate}%
        \hskip -\textwidth \hbox to \textwidth{\hfil\rm\thepage\hfil}}%
     \else\def\@oddfoot{\hbox to \textwidth{\hfil\rm\thepage\hfil}}
     \fi
     \let\@evenfoot\@oddfoot
}

\def\@version#1{\ifnum\draftcontrol=1
\typeout{}\typeout{#1}\typeout{}
\vskip3mm\centerline{\hbox{\fbox{\normalsize{\tt DRAFT -- #1 -- }
                   {\draftdate}}}}\vskip3mm
\fi}
\let\version\@version
\long\def\eqlabel#1{\ifnum\draftcontrol=1
                    \tag@false  
                    \tag*{(\theequation) \hbox to -0.2cm{\hspace{0cm}\small{#1}\hss}}
                    \refstepcounter{equation}
                    \edef\@currentlabel{\theequation}
                    \ltx@label{#1}          
                    \else
                    \label{#1}
                    \fi
                    }
\let\st@bibitem\@bibitem
\let\st@lbibitem\@lbibitem
\ifnum\draftcontrol=1
  \def\@bibitem#1{%
    \st@bibitem{#1}\a@@label{#1}\ignorespaces}
  \def\@lbibitem[#1]#2{%
    \st@lbibitem[#1]{#2}\a@@label{#2}\ignorespaces}
  \def\a@@label#1{%
    \gdef\a@lab{\smash{\normalfont\small#1}}
    \ifvmode
      \if@inlabel
        \global\setbox\@labels\hbox{%
          \llap{\a@lab\let\a@lab\relax
                \kern\@totalleftmargin\kern\marginparsep}%
          \box\@labels}%
      \fi
    \fi}
\fi

\documentclass[12pt,letterpaper]{article}

\usepackage{amsmath,amssymb,array,calc,rotating,epsfig,latexsym,psfrag}
\usepackage[nosort]{cite}
\ifnum\draftcontrol=1
\fi

\renewcommand\baselinestretch{1.25}
\renewcommand\section{\@startsection {section}{1}{\z@}%
                                   {-3.5ex \@plus -1ex \@minus -.2ex}%
                                   {2.3ex \@plus.2ex}%
                                   {\normalfont\large\bfseries}}
\renewcommand\subsection{\@startsection{subsection}{2}{\z@}%
                                   {-3.25ex\@plus -1ex \@minus -.2ex}%
                                   {1.5ex \@plus .2ex}%
                                   {\normalfont\normalsize\bfseries}}
\renewcommand\subsubsection{\@startsection{subsubsection}{3}{\z@}%
                                   {-3.25ex\@plus -1ex \@minus -.2ex}%
                                   {1.5ex \@plus .2ex}%
                                   {\normalfont\normalsize\it}}
\renewcommand\paragraph{\@startsection{paragraph}{4}{\z@}%
                                   {-3.25ex\@plus -1ex \@minus -.2ex}%
                                   {1.5ex \@plus .2ex}%
                                   {\normalfont\normalsize\bf}}

\def\revise#1       {\raisebox{-0em}{\rule{3pt}{1em}}%
                     \marginpar{\raisebox{.5em}{\vrule width3pt\
                     \vrule width0pt height 0pt depth0.5em \hbox to
                     0cm{\hspace{0cm}{%
                     \parbox[t]{4em}{\raggedright\footnotesize{#1}}}\hss}}}}

\def\del          {\partial}

\def\tr           {\mathop{\rm Tr}}

\def\half{{\frac12}}

\def\sqr#1#2{{\vcenter{\vbox{\hrule height.#2pt
 \hbox{\vrule width.#2pt height#1pt \kern#1pt \vrule width.#2pt}\hrule
 height.#2pt}}}}

\def\a{\alpha}
\def\b{\beta}
\def\r{\rho}

\def\g{\gamma}

\catcode`\@=12

\begin{document}


\topmargin=0.50in

\newcommand{\be}{\begin{equation}}
\newcommand{\ee}{\end{equation}}
\newcommand{\beq}{\begin{equation}}
\newcommand{\eeq}{\end{equation}}
\newcommand{\ba}{\begin{eqnarray}}
\newcommand{\ea}{\end{eqnarray}}
\newcommand{\nn}{\nonumber}

\def\vol{\bf vol}
\def\Vol{\bf Vol}
\def\del{{\partial}}
\def\vev#1{\left\langle #1 \right\rangle}
\def\cn{{\cal N}}
\def\co{{\cal O}}
\def\IC{{\mathbb C}}
\def\IR{{\mathbb R}}
\def\IZ{{\mathbb Z}}
\def\RP{{\bf RP}}
\def\CP{{\bf CP}}
\def\Poincare{{Poincar\'e }}
\def\tr{{\rm tr}}
\def\tp{{\tilde \Phi}}
\def\Y{{\bf Y}}
\def\te{\theta}
\def\bX{\bf{X}}

\def\TL{\hfil$\displaystyle{##}$}
\def\TR{$\displaystyle{{}##}$\hfil}
\def\TC{\hfil$\displaystyle{##}$\hfil}
\def\TT{\hbox{##}}
\def\HLINE{\noalign{\vskip1\jot}\hline\noalign{\vskip1\jot}} 
\def\seqalign#1#2{\vcenter{\openup1\jot
  \halign{\strut #1\cr #2 \cr}}}
\def\lbldef#1#2{\expandafter\gdef\csname #1\endcsname {#2}}
\def\eqn#1#2{\lbldef{#1}{(\ref{#1})}%
\begin{equation} #2 \label{#1} \end{equation}}
\def\eqalign#1{\vcenter{\openup1\jot
    \halign{\strut\span\TL & \span\TR\cr #1 \cr }}}
\def\eno#1{(\ref{#1})}
\def\href#1#2{#2}
\def\half{{1 \over 2}}

\def\ads{{\it AdS}}
\def\adsp{{\it AdS}$_{p+2}$}
\def\cft{{\it CFT}}

\newcommand{\ber}{\begin{eqnarray}}
\newcommand{\eer}{\end{eqnarray}}

\newcommand{\bea}{\begin{eqnarray}}
\newcommand{\eea}{\end{eqnarray}}

\newcommand{\beqar}{\begin{eqnarray}}
\newcommand{\cN}{{\cal N}}
\newcommand{\cO}{{\cal O}}
\newcommand{\cA}{{\cal A}}
\newcommand{\cT}{{\cal T}}
\newcommand{\cF}{{\cal F}}
\newcommand{\cC}{{\cal C}}
\newcommand{\cR}{{\cal R}}
\newcommand{\cW}{{\cal W}}
\newcommand{\eeqar}{\end{eqnarray}}
\newcommand{\lm}{\lambda}\newcommand{\Lm}{\Lambda}
\newcommand{\eps}{\epsilon}


\newcommand{\nonu}{\nonumber}
\newcommand{\oh}{\displaystyle{\frac{1}{2}}}
\newcommand{\dsl}
  {\kern.06em\hbox{\raise.15ex\hbox{$/$}\kern-.56em\hbox{$\partial$}}}
\newcommand{\as}{\not\!\! A}
\newcommand{\ps}{\not\! p}
\newcommand{\ks}{\not\! k}
\newcommand{\D}{{\cal{D}}}
\newcommand{\dv}{d^2x}
\newcommand{\Z}{{\cal Z}}
\newcommand{\N}{{\cal N}}
\newcommand{\Dsl}{\not\!\! D}
\newcommand{\Bsl}{\not\!\! B}
\newcommand{\Psl}{\not\!\! P}
\newcommand{\eeqarr}{\end{eqnarray}}
\newcommand{\ZZ}{{\rm \kern 0.275em Z \kern -0.92em Z}\;}

\def\s{\sigma}
\def\a{\alpha}
\def\b{\beta}
\def\r{\rho}
\def\d{\delta}
\def\g{\gamma}
\def\G{\Gamma}
\def\ep{\epsilon}
\makeatletter \@addtoreset{equation}{section} \makeatother
\renewcommand{\theequation}{\thesection.\arabic{equation}}

\begin{titlepage}

\version\versionno

\leftline{\tt hep-th/0401031}

\vskip -1cm

\rightline{\small{\tt MCTP-03-62}}
\rightline{\small{\tt CPHT-RR-114-1203}}
\rightline{\small{\tt IC/2003/162}}
\rightline{\small{\tt MIT-CTP 3455}}


\vskip .7 cm

\centerline{\bf \Large On the Universality Class of Certain String Theory Hadrons }

\vskip .8cm
{\large }

\centerline{\large  G. Bertoldi${}^1$, F. Bigazzi${}^2$, A. L.
Cotrone${}^{3}$, C. N\'u\~nez${}^4$, L.
A. Pando Zayas${}^{5}$}

\vskip .5cm \centerline{\it ${}^1$ School of Natural Sciences, Institute for Advanced Study }
\centerline{ \it Einstein Drive,  Princeton, NJ 08540. USA}

\vskip .2cm \centerline{\it ${}^2$ The Abdus Salam ICTP,
Strada Costiera, 11; I-34014 Trieste, Italy}

\vskip .2cm \centerline{\it ${}^3$ CPHT,  \'Ecole Polytechnique,
F-91128 Palaiseau Cedex, France}

\vskip .2cm \centerline{\it ${}^3$INFN,  Piazza dei Caprettari, 70; I-00186  Roma, Italy}

\vskip .2cm \centerline{\it ${}^4$ Center for Theoretical Physics, MIT,
Cambridge MA02140. USA}

\vskip .2cm \centerline{\it ${}^5$ Michigan Center for Theoretical
Physics} \centerline{ \it Randall Laboratory of Physics, The
University of Michigan} \centerline{\it Ann Arbor, MI 48109-1120.
USA}

\vspace{.5cm}

\begin{abstract}
Exploiting the gauge/gravity correspondence we find the spectrum of
hadronic-like bound states of adjoint particles with a large global
charge in several confining
theories. In particular, we consider an embedding
of four-dimensional ${\cal N}=1$ supersymmetric
Yang-Mills into IIA string theory, two classes of three-dimensional gauge theories and the softly broken version of one of them.
In all cases  we describe the low energy
excitations of a heavy hadron with mass proportional to its global
charge. These excitations include: the hadron's nonrelativistic motion,
its stringy
excitations and excitations corresponding to the addition of massive
constituents.
Our analysis
provides ample evidence for the universality of such hadronic states in
confining theories admitting supergravity duals.
Besides, we find numerically a new smooth solution that can be thought
of as  a non-supersymmetric deformation of $G_2$ holonomy manifolds.
\end{abstract}



\end{titlepage}

\newpage

\tableofcontents

\pagebreak


\section{Introduction and Summary}
The gauge/gravity correspondence has reached a new chapter
with the realization of concrete scenarios where the correspondence
can be taken beyond the supergravity approximation \cite{bmn,gkp}. An
interesting example was given by Berenstein, Maldacena and Nastase who
proposed a gauge theory interpretation of the Penrose-G\"uven limit
for IIB supergravity on $AdS_5\times S^5$. A crucial ingredient in the
BMN construction is that string theory in the resulting background is
exactly soluble \cite{metsaev}.

Inspired by this new insight, the authors of reference \cite{gpss}
considered a modification of the Penrose limit applicable to
supergravity backgrounds dual to confining gauge theories.
By focusing in on the IR
region, exactly solvable string theory models were obtained. These
string  models represent in the gauge theory side the nonrelativistic motion and low-lying excitations
of heavy hadrons with mass proportional to a large $U(1)$  global
charge.

The hadronic states discussed in \cite{gpss} are remarkable in  that
they are present in any confining theory admitting a supergravity
dual. More precisely, as shown in \cite{gpss}, the conditions for the
existence of a Penrose-G\"uven limit that focuses in on these states
is compatible with the condition for a supergravity background to
allow an area law for the VEV of a rectangular 
Wilson loop in the dual
gauge theory.

In \cite{gpss} the authors considered the Klebanov-Strassler (KS) \cite{ks}
and the Maldacena-N\'u\~nez  (MN) \cite{mn} backgrounds, both  embeddings
of 4-d ${\cal N}=1$  SYM into IIB string theory. The corresponding
confining gauge theories consist of ${\cal N}=1$ SYM plus massive
particles in the adjoint representation carrying a global $U(1)$
charge. After taking the Penrose limit it was realized that the string Hamiltonian on the resulting background describes stringy shaped hadrons, called annulons, which are bound states of
these massive particles, in the limit where the charge and the number
of colors both go to infinity.

The analysis of \cite{gpss}  attempts to
elucidate which properties of the annulons  are universal to any
background dual to a confining theory and which depend on the
particular embedding.  Recently, softly broken versions of the KS
and MN  backgrounds were  considered in \cite{alfra,Kuperstein:2003yt}
with results very
similar to those of
\cite{gpss}.

In this paper we explore other string duals of confining gauge theories,
and show how those universal properties are in fact manifest.

The first model we consider is an embedding of 4-d ${\cal N}=1$ SYM into
IIA string theory.  The field theory lives on the worldvolume of $N$
D6-branes wrapped on a three-cycle of a deformed conifold.  The dual IIA
background results in a product of a flat 4-d Minkowski space and a
resolved conifold, and it is supported by $N$ units of RR two-form flux
through the non vanishing two-cycle.  This 10-d solution is obtained as
a $U(1)$ compactification of an M-theory background with an
asymptotically locally conical, $G_2$ holonomy metric \cite{iiacv,iiab}.
The circle along which we compactify has finite size everywhere; this
ensures that the  10-d solution is regular.


As a second class of models, we consider IIB and IIA string duals to
three dimensional ${\cal N}=1$ confining gauge theories.  The relevant
backgrounds were given in
\cite{Chamseddine:2001hk,Schvellinger:2001ib,maldanastase} and
\cite{CGLP} respectively.  In a sense they are the 3-d analogue of the
MN and KS backgrounds. The solution in \cite{maldanastase} describes the
geometry generated by fivebranes wrapped over a three-cycle of a
(topologically) $G_2$ holonomy manifold; recall that the MN background
describes D5-branes wrapped on a two-cycle.  The one in \cite{CGLP}
corresponds to a stack of regular and fractional D2-branes whereas KS
corresponds to a stack of D3 and fractional D3-branes.  We also
consider the gluino-mass-deformed version of the gauge model dual to the
solution in \cite{Chamseddine:2001hk,Schvellinger:2001ib,maldanastase}.
The corresponding non supersymmetric background was studied in
\cite{gae}.

The understanding of 3-d gauge theories in the context of the AdS/CFT correspondence is less precise.
In fact, the AdS/CFT correspondence has been used to predict
various properties of confining 3-d gauge theories \cite{herzog} \footnote{An example of a
property of 4-d confining gauge theories that has been understood in
the gauge/gravity correspondence and whose derivation has
been used to conjecture the behavior of confining 3-d gauge theories
is the tension of a string ending on $q$ external quarks.}.
We thus present our analysis as an alternative way of gaining information
about 3-d confining  gauge theories. Having in some sense established
the universality of certain states in embeddings of 4-d  confining
theories we proceed to study the corresponding states in 2+1
dimensions.

Finally, with the intention of exploiting a non supersymmetric version
of the 4-d model above, we numerically find a new set of solutions that
are non supersymmetric deformations of $G_2$ holonomy manifolds. These
solutions are smooth and, depending on the range of  parameters, we can
attempt an interpretation as duals to D6-branes  wrapping  cycles in a
non-supersymmetric fashion. The brane interpretation suggests that these
solutions might be considered as new gravity duals 
of confining non supersymmetric gauge theories, though
more work is needed to precisely state this conjecture.  As a first
plausibility indication we analyze the Penrose limit of the new
backgrounds, finding a pp-wave solution identical to the one found in
the susy case: this should give us some evidence about the presence of
annulons also in this nonsupersymmetric context.

The paper is organized as follows. In the following section we give a
general overview of the problem as well as a description of the main
principle behind our analysis. Section \ref{iia} contains a description
of our Penrose limit applied to the resolved conifold with RR two-form
and the spectrum of IIA string theory on the resulting background. In
section \ref{maldana} we describe the Penrose limit of the D5-on-$S^3$
supersymmetric and softly broken solutions.  Section \ref{d2fd2} deals
with the D2 and fractional D2-branes  model. Section \ref{gaugetheory}
contains our analysis of the string theory results from the gauge theory
point of view. We identify the constituents of the ground state hadrons
and the corresponding excitations. In section \ref{nonsusyiia} we
present the nonsupersymmetric deformation of the background studied in
section \ref{iia}.
Section \ref{conc} contains some concluding remarks.
We have also included a number of appendices that contain explicit
calculations of technical statements made in the body of the
paper. Appendix \ref{appendixa} contains standard expressions for the
$SU(2)$ left-invariant forms and a parametrization of $\mathbb{R}^4$
used in the main body. Appendix \ref{appendixb} contains the fermionic
string equations of motion for a generic plane wave background. The
BPST instanton needed for the explicit form of the solution treated in 
section
\ref{d2fd2} is given explicitly in appendix \ref{appendixc}. Appendix
\ref{zeropoint} contains a discussion of the regularization used to
compute the zero point energy of the string theories obtained in the
main body. Appendix \ref{appendixd} presents the full set of second
order differential equations  that are solved in section
\ref{nonsusyiia} to obtain the new nonsupersymmetric deformation of the $G_2$
holonomy metric. This appendix also contain fifteen figures designed
to present a graphical proof of the  existence and properties of
the new solution.

{\bf Reader's guide:} Since this is a long paper, it is useful to give
a `road-book' for readers with diverse interests. Those interested in the `geometrical'
aspects of the paper, like Penrose limits, pp-waves and non supersymmetric deformations of $G_2$ holonomy manifolds,
should read sections 3.1, 4, 5.1, and section 7 (this last one can be
read almost independently of the rest
of the paper). Readers more concerned with string theory aspects of this work, should refer to
sections 3.2, 3.3, 4.1, 5.2, 6
and appendix D.

\section{Physical picture of the general idea}
One of the main purpose of this paper is to address the universality
of the annulons, i.e. those hadronic states first considered in \cite{gpss}.

A controlled setup in which to study gauge theory IR dynamics,
including confinement, would be in the presence of supersymmetry.
Four-dimensional ${\cal N}=1$ SYM provides such a system
\footnote{Nevertheless, what we say in this section also applies to the
three-dimensional and non-supersymmetric theories.}. In the
context of the gauge/gravity correspondence this would correspond
to studying the string theory dual to ${\cal N}=1$ SYM which is
not known. The best we can do at the moment is to study
supergravity backgrounds that correspond to confining gauge
theories that contain ${\cal N}=1$ SYM as a sector. In all known
cases these confining gauge theories contain massive fields that
transform in the adjoint of the gauge group and have some global
charge.  If we group enough of these massive particles we might
hope that they  form bound states. This is precisely the field
theoretic question we address: {\it What is the spectrum of a
collection of a large number of these massive, globally charged,
adjoint particles?}

The question just formulated about the spectrum of a bound state of
adjoint particles with global charge is out of the reach of current
field theoretic techniques. However, the gauge/gravity correspondence
provides an {\it exact} answer.

The general principle allowing to address such questions in string
theory was articulated by  Gubser, Klebanov and Polyakov in
\cite{gkp}.  In the gauge/gravity correspondence there is a direct
relation between sectors of the gauge theory with large quantum
numbers and classical solitonic solutions of the string sigma model on
the corresponding supergravity background. Under this point of view
the BMN sector of operators \cite{bmn} corresponds to a string shrunk
to a point and orbiting at the speed of light along the great circle
of $S^5$. For BMN operators in the gauge theory, the R-charge is to be
identified with  the angular momentum of the classical
solution. Applying this principle to the question at hand we can
translate the spectrum of a bound state of a large number of globally
charged massive
adjoint particles into the semiclassical analysis of a macroscopic
string that spins in the space perpendicular to the worldvolume of the
gauge theory. The angular momentum of the classical string is to be
identified with the global charge of the bound state. As shown in
\cite{gpss}, such a string configuration can only be stationary  in the
region corresponding to the IR of the gauge theory.  Thus the
classical configuration we consider is a string stuck at the ``minimal
radius''\footnote{This name stems from the fact that in $AdS$ in 
Poincare coordinates the
space ends at $r=0$ which corresponds to  $g_{00}=0$. In supergravity
theories dual to confining backgrounds the spaces ends at a minimal
value of $r$ for which $g_{00}\ne 0$ \cite{cobicond}, this value $r_0$ is called the
minimal AdS radius.}   and spinning  in the internal space.

It is worth pointing out the role of this region for the states we
describe. A hadron, being a gauge theory state of definite
four-dimensional mass, is dual to a supergravity eigenstate of the
ten- and four-dimensional Laplacians. In the world volume directions
these states are plane waves. The wave function in the remaining
directions falls off as $\psi(r,\Omega)\sim r^{-\Delta}$, where
$\Delta$ is the dimension of the lowest-dimension operator which can
create the hadron. Therefore, a hadron made of a large number of
constituents is localized near the ``minimal AdS radius.''

It is remarkable that for this particular configuration we can provide
an {\it exact} analysis. The reason is that it can be studied by means
of a particular Penrose limit. This effectively implies, as in the BMN
case, that the semiclassical quantization is exact. The Penrose limit
provides a truncation of the supergravity background to one in which
the corresponding string theory can be exactly solved.


\section{Resolved conifold with RR two-form flux}
\label{iia}
In this section we  consider a IIA background argued to be  dual to an
embedding of ${\cal N}=1$ SYM into string theory.  This background is
a resolved conifold metric with RR two-form flux turned on over the
blown up $S^2$.

The simplest interpretation of this background is as originating from
a  $G_2$  holonomy metric in the $D_7$ family \cite{iiacv,iiab}.
We will use this family because of its good short distance behavior that
translates into the desirable IR properties of the dual gauge theory.
In
eleven dimensions the background is just a metric of the form
\begin{equation}
\eqlabel{gg2}
\begin{split}
ds_{11}^2 &= dx_{1,3}^2 + ds^2_{G_2},  \\ ds^2_{G_2}&=dr^2 + a(r)^2
[(\Sigma_1 + g(r) \sigma_1)^2 +  (\Sigma_2 + g(r) \sigma_2)^2]\\ &+
c(r)^2 (\Sigma_3 + g_3(r) \sigma_3)^2 +  b(r)^2 (\sigma_1^2 +
\sigma_2^2) + f(r)^2 \sigma_3^2,
\end{split}
\end{equation}
with $g(r) =-\frac{a f}{2 b c},\;\;  g_3(r) = 2 g(r)^2 -1$. As usual
$\Sigma, \sigma$ are left invariant forms on each one of the
$SU(2)'s$ of the symmetry group of the $G_2$ manifold.  See appendix
\ref{appendixa} for the conventions.

Due to the present  number of $U(1)$ isometries, one can reduce this
metric to type IIA. In order for the ten dimensional background to be
a good supergravity background, one must  impose that asymptotically
(for large values of the radial coordinate),  there is a stabilized
one cycle. In other words, we require the metric to be Asymptotically
Locally Conical (ALC).

None of the radial functions is known explicitly, although the
asymptotics at the origin and at infinity are known.  The
equations \footnote{We follow the notation of \cite{iiacv}. However, we
refer the reader to \cite{iiab} for a more comprehensive analysis of
the general properties of the solution. In particular, the large $r$
asymptotic is more exhaustive.} are \cite{iiacv}

\bea
\label{eq:Deqns}
\dot{a} = -\frac{c}{2a} + \frac{a^5 f^2}{8 b^4 c^3}, & \;\; &
\dot{b} = -\frac{c}{2b} - \frac{a^2 (a^2-3c^2)f^2}{8b^3c^3}, \nn \\
\dot{c} = -1+\frac{c^2}{2 a^2}+\frac{c^2}{2 b^2}-\frac{3 a^2
f^2}{8b^4}, & \;\; &
\dot{f} = -\frac{a^4 f^3}{4 b^4 c^3}.
\eea
As $r\to 0$ one has
\begin{equation}
\eqlabel{r0}
\begin{split}
a(r) & =  \frac{r}{2}-\frac{(q_0^2+2)r^3}{288 R_0^2} -
\frac{(-74-29q_0^2+31q_0^4)r^5}{69120 R_0^4} + \cdots  ,  \\
b(r) & =  R_0 - \frac{(q_0^2-2)r^2}{16 R_0} -
\frac{(13-21q_0^2+11q_0^4) r^4}{1152 R_0^3} + \cdots  , \\
c(r) & =  -\frac{r}{2} - \frac{(5q_0^2-8)r^3}{288 R_0^2} -
\frac{(232-353q_0^2+157q_0^4) r^5}{34560 R_0^4}+ \cdots ,  \\
f(r) & =  q_0 R_0 + \frac{q_0^3 r^2}{16 R_0} +
\frac{q_0^3(-14+11q_0^2) r^4}{1152 R_0^3} + \cdots ,
\end{split}
\end{equation}
where $q_0$ and $R_0$ are constants. In \cite{iiacv,iiab} it was
pointed out through a numerical analysis that  the solution is ALC and
regular if  $q_0\in(0,1)$.  Note that $a(r)$ and $c(r)$ vanish and the
other two functions do not.  As $r\to\infty$ we have
\bea\label{eq:Datinfinity}
a(r) & = &  \frac{r}{\sqrt{6}} - \frac{\sqrt{3} q_1 R_1}{\sqrt{2}} +
\frac{(27\sqrt{6}- 96 h_1 ) R_1^2}{96r} + \cdots , \nn \\
b(r) & = &  \frac{r}{\sqrt{6}} - \frac{\sqrt{3} q_1 R_1}{\sqrt{2}} +
\frac{h_1 R_1^2}{r} + \cdots , \nn \\
c(r) & = &  \frac{-r}{3} + q_1 R_1 - \frac{9 R_1^2}{8r} + \cdots , \nn
\\ f(r) & = &  R_1 - \frac{27 R_1^3}{8 r^2} - \frac{81 R_1^4 q_1}{4
r^3} +
\cdots ,
\eea
with constants $R_1,q_1,h_1$\footnote{They should be related to the
small $r$ expansion constants by using the fact that the resolved
conifold parameter $p=f(a^4f^2+4b^4c^2-4a^2b^2c^2)/(4b^2c^2)$ has to
be a constant \cite{Cvetic:2001kp}.}. Note that $f(r)$ stabilizes,
explicitly realizing the ALC condition.

Three constants appear to this order, whilst there were only two
constants in the expansion about the origin. This just means that for
some values of these constants, the corresponding solution will
diverge before it reaches zero. In any case, we find no $h_1$
dependence in the results below.

There are two natural $U(1)$ isometries to be considered in reducing
to type IIA; these are linear combinations of the two angles ($\psi,
\bar{\psi}$)  that appear in the left invariant forms ($\sigma,
\Sigma$) respectively. The convenient linear combinations are
$\psi_2 = \bar{\psi} - \psi$ and $\psi_1 = \bar{\psi} + \psi$ (the
combination $\psi_2$ appears in the metric and background fields in
IIA  while the other does not).

In order to obtain a IIA background involving $N$ units of RR two-form
flux trough the non collapsing two-sphere we have to mod the 11-d
metric by $\psi_1\rightarrow \psi_1+4\pi/N$. We refer to
\cite{maldavafa} for the details of the calculation in an analogous
case. The IIA background metric reads (in string frame)

\bea
\label{metriiia}
&&e^{-2\Phi/3} ds_{10}^2 = dx_{1,3}^2 + \ell^2\bigl[ dr^2 + a(r)^2 (d\bar{\theta}^2 +
\sin^2\bar{\theta}d\bar{\phi}^2)+\nonumber \\
&& + (b(r)^2 +
a(r)^2g(r)^2)(d\theta^2 +
\sin^2\theta d\phi^2)+ \\
&&+\frac{f^2 c^2}{f(r)^2 + (1 + g_3(r))^2
c(r)^2}
(d\psi_2 -
\cos\theta d\phi + \cos\bar{\theta} d\bar{\phi})^2+ \nonumber \\
&&+ 2 g(r) a(r)^2
[\cos\psi_2 (d\theta d\bar{\theta} + \sin\theta \sin\bar{\theta}d\phi
d\bar{\phi}) + \sin\psi_2 (\sin\bar{\theta} d\theta
d\bar{\phi} - \sin\theta d\phi
d\bar{\theta})]\bigr]\nonumber,
\eea
and the matter fields are given by
\bea
\label{matteriia}
&&A_1= \ell N\bigg[\frac{f(r)^2 - (1
- g_3(r)^2) c^2}{f(r)^2 + (1 +
g_3(r))^2 c(r)^2} (-d\psi_2 +
\cos\theta d\phi - \cos\bar{\theta} d\bar{\phi}) + \nonumber \\
&& \qquad \qquad +\cos\theta d\phi +
\cos\bar{\theta} d\bar{\phi}\bigg],\nonumber \\
&&4N^2e^{4\Phi/3}= f(r)^2 + c(r)^2 (1 + g_3(r))^2,
\eea
where we have introduced a natural dimensionful parameter $\ell\approx\sqrt{\alpha'}$.

According to the expansions above, one of the two-cycles, the one given
by the barred coordinates ($\bar{\theta}, \bar{\phi}$), shrinks to zero
size near $r=0$, while the
other does not, since the function $b(r)^2 + [a(r)g(r)]^2$ does not
vanish near $r=0$. Typically, in these backgrounds, there
is a fibration between both cycles. Note that we
now get $N$ units of flux for $F_{(2)}=dA_1$ through the non collapsing sphere at infinity.
The non-vanishing cycle is a natural candidate for the $U(1)$ direction
needed to perform the Penrose-G\"uven limit which we discuss in the next
subsection.

Let us now discuss some aspects of the gauge/gravity duality in this case.
A simple way to understand the solution above is to turn
to its M-theory interpretation. At the eleven dimensional level, this
background which is purely metric, is dual at low energies to ${\cal N}=1$
SYM with $SU(N)$ gauge group coupled to additional massive KK modes.
The supergravity solution encodes the information as
follows. The amount of supersymmetry is a consequence of
the metric exhibiting $G_2$ holonomy. The rank of the gauge group is the
result of modding out the compactifying $U(1)$ by ${\mathbb Z}_N$.
The duality proposed by Vafa in
\cite{Vafa:2000wi}, stating that the gauge theory obtained by wrapping D6 branes
on a three cycle of the deformed conifold is dual to the background
written in (\ref{metriiia})-(\ref{matteriia}), was put in M-theory perspective by
\cite{maldavafa} as a topology change, thus starting the study of non compact
$G_2$ holonomy manifolds
in the context of dualities between gauge theories and M-theory backgrounds.
Quantum aspects of this correspondence have been further studied in
\cite{Atiyah:2001qf}. The connection of ${\cal N}=1$ SYM with the $ADE$ structure of the
singularities of these manifolds was explained in
\cite{Acharya:2000gb}-\cite{Acharya:gu}.
One should keep in mind that these developments
deal mainly with topological properties of $G_2$ holonomy manifolds,
 the explicit form of the metrics being
almost of no relevance. Indeed, various topological aspects,
related to
the confining strings of ${\cal N}=1$ SYM appearing as M2-branes wrapping a one-cycle, or domain
walls, as M5-branes wrapping three cycles in the geometry, have been studied in
\cite{Acharya:2001hq,Acharya:2001dz}. It is then of interest
to gain insight on gauge theory quantities that do depend explicitly
on the form of the metric. From the gauge theory point of view these
quantities are generically more dynamical. There are
few examples of this type of calculations.
One can mention the study of rotating membranes,
that are argued to be dual to large spin operators in ${\cal N}=1$
SYM, reproducing the well known relation between the spin and the
energy of the state  (wrapped membrane) $E- S = Log[S]$,
\cite{Hartnoll:2002th}. Another nice example of a
dynamical  gauge theory quantity that depends
explicitly on the form of the metric is given by the
study of the chiral anomaly of the R-symmetry of ${\cal N}=1$  SYM. We refer the reader to
\cite{Gursoy:2003hf} for a neat study of this subtle point.
These authors also signaled that the understanding of the breaking
$U(1)\to {\mathbb Z}_{2N}$ might require the construction of a new background, that should share
some of the features of the ones already known.

The problem we deal with in this paper belongs to the class of
dynamical questions mentioned above,
that is, its result will depend on the form of the $G_2$ manifold we use.
Indeed,  the study of (adjoint)
hadrons, composed out of a large number of massive excitations of the field theory
(KK modes), its spectrum and interactions is a dynamical problem of interest, that involves crucially
particular aspects of the metric mentioned above. Then, we see the content of the next subsection as a
nice `dynamical experiment'
with $G_2$ holonomy manifolds as duals to gauge theories with minimal supersymmetry in four dimensions.

The study of the gauge/gravity correspondence in this IIA context
is not transparent since the resource of deforming the
original AdS/CFT correspondence is not available. There are many
features that are not fully understood. Let us comment on this.
In other set ups that are not asymptotically $AdS$, like for example the KS case \cite{ks},
one can advocate the fact that for large radius
\footnote{The logarithmic behavior is a property exclusively related to the UV completion of various  gauge theories
\cite{kt,pt}.} the background can be roughly seen as a ``logarithmic'' deformation of $AdS$,
and so try to implement the standard (UV) relation $r\approx \mu/\Lambda$ with the field theory scale.
This identification turns out to be correct  {\it a posteriori}, and in fact gives a
correct prediction for the logarithmic running of the dual gauge theory beta
functions \cite{kt,ks}, and it is consistent with the expected
scaling of the duals of the gaugino condensates \cite{bgz,ls}.

For the background in \cite{mn}, a suggestion for the explicit radius/energy relation, at
least in the UV, was given in \cite{abcpz}. This
background describes the strong coupling regime of a stack of IIB fivebranes wrapping a
two-cycle inside the resolved conifold.
Then, by identifying the supergravity dual
of the gluino condensate,
one can compute the running of the gauge coupling in the geometric
dual, reproducing
the well known NSVZ result
\cite{abcpz,DiVecchia:2002ks,Bertolini:2002yr}.

Several things need to be improved in the context of
this  M-theory/gauge theory duality, based on $G_2$ holonomy
manifolds. Some of them include the precise identification of the
radius/energy relation, the geometric dual of the gluino condensate, the
cycle that
the D-branes are wrapping \footnote{See \cite{Edelstein:2002zy} for a  possible resolution of this
issue.},
the definition of the gauge coupling and its running, the breaking of the global symmetry
$U(1)_R \to {\mathbb Z}_{2N}$ and also the non-decoupling of the KK
modes. Despite some effort these topics are still unclear. Perhaps a
more complete understanding requires   a  not yet known supergravity solution.

After summarizing the points that are not clear with this duality,
we can proceed in analogy with the case of flat D6-branes, where
large radial distances represent the UV of the seven dimensional
gauge theory living on the brane stack. Along this line, when we
wrap the branes on the three-cycle we get a solution that we
interpret in the same way. The large values of the radial
coordinate will be dual to the UV region of the four dimensional
gauge theory (${\cal N}=1$  SYM plus KK modes), while  the IR will
be encoded in small values of the radial coordinate. We emphasize
that there are many  indications that this identification is
correct. Indeed, for the $D_7$ family of metrics (the one we deal
with in here), the short distance expression of (\ref{metriiia})
implies confinement in the dual gauge theory and also a breaking
of the global symmetry to ${\mathbb Z}_2$. The hadrons that we will work
with in this section are purely IR effects as will be shown below.

\subsection{A Penrose limit}

We thus consider taking a  Penrose limit that naturally zooms in on the
region of small values of the radial coordinate $r$. The way of doing
this in confining backgrounds was explained in \cite{gpss,alfra}.
Though not strictly necessary, we want to take
the Penrose limit in  a way that the scale parameter which is
eventually taken to infinity is linked to some physical quantities of our background.
But, on the other hand, in the cases considered in
this paper the dilaton is not constant and we have to ensure that in
the limit we do not end up with an infinite value for it.
Thus, let us first rescale the flat 4-d coordinates
by means of $x^{\mu}\rightarrow Le^{-\Phi_0/3}x^{\mu}$,
where $\Phi_0$ is the value of the dilaton at the origin.
This way our metric in the extreme IR will formally read
\begin{equation}
ds^2\approx 2\pi\alpha' T_sdx^2_{1,3} + {2\pi\alpha'T_s\over m_0^2}ds_6^2
\end{equation}
where, in the particular case at hand,
\begin{equation}
\label{stringscale}
T_s = {L^2\over{2\pi\alpha'}}, \qquad m_0^2= {L^2\over{\ell^2R_0^2e^{2\Phi_0/3}}}.
\end{equation}
The limit we will take \cite{gpss} will send $T_s$ to infinity while
keeping  $m_0$ fixed: this amounts to taking $L^2\approx R_0^2e^{2\Phi_0/3}\to\infty$
\footnote{Note that we have $2e^{2\Phi_0/3}= q_0R_0/N$. If we want that $L\to\infty$ corresponds to $N\to\infty$ as in all the other
cases we examine, we have to take $R_0$ dependent on $N$. One possibility is to take the dilaton
constant while performing the $N\to\infty$ limit. This would require $R_0\approx N$ and so $L^2\approx N^2$.}.
Note that in the IIB context \cite{gpss}, $T_s$ and $m_0$ are related to 
the confining strings tension and KK (or glueball) masses. Here we adopt the same notations
though, perhaps,  the relation with gauge theory is more subtle.

From the form of the radial functions in the IR (\ref{r0}) it can be deduced that the natural coordinate with a $U(1)$ symmetry along which we could
take a  Penrose limit is $\phi$. We therefore expand around
\begin{equation}
\eqlabel{expansion}
r\to {m_0R_0r\over L}, \qquad \theta \to {\pi\over 2}-{m_0v\over L}, \qquad
x_i\to {x_i\over L},\,\, i=1,2,3.
\end{equation}
Expanding the metric up to quadratic terms in $L^{-1}$ we obtain
\begin{equation}
\eqlabel{quad}
\begin{split}
ds^2&=(1+{3q_0^2\over{32}} {m_0^2r^2\over L^2})(-L^2dt^2+ {L^2\over m_0^2}d\phi^2)+ dx^idx_i + dr^2+\\
&+{r^2\over 4 }[d\bar\theta^2+\sin^2\bar\theta d\bar\phi^2
+(d\psi_2 +\cos\bar\theta d\bar\phi)^2]+dv^2- [v^2+{(q_0^2-4)r^2\over16}]d\phi^2 +\\
& + {q_0r^2\over4}(\cos\psi_2\sin\bar\theta d\bar\phi - \sin\psi_2d\bar\theta)d\phi.
\end{split}
\end{equation}
We will then pass to the light-cone coordinates
\begin{equation}
\eqlabel{lccoordinates}
\quad x^+=t, \qquad x^- = {L^2\over2}(t-{\phi\over m_0}).
\end{equation}
Before writing out the final result, let us consider the four
directions $(r,\bar\theta,\psi_2,\bar\phi)$. It can be seen that they
parametrize an $\mathbb{R}^4$. We  parametrize this $\mathbb{R}^4$
by four Cartesian coordinates $(y_1,y_2,y_3,y_4)$. The term that mixes this
coordinates and the $U(1)$ direction $\phi$ is, in the notation of
appendix \ref{appendixa},  simply
\begin{equation}
{1\over 2}r^2\sigma_2=y_1dy_3-y_3dy_1+y_2dy_4-y_4dy_2.
\end{equation}
With all  substitutions included, the metric takes  the form
\begin{equation}
\eqlabel{metric}
\begin{split}
ds^2&=-4dx^+dx^- - m_0^2\bigg[v^2+{(q_0^2-4)\over16}y^ay_a\bigg](dx^+)^2+dx^idx_i+dy^ady_a+dv^2 \\
&+{m_0q_0\over 2}dx^+(y_1dy_3-y_3dy_1+y_2dy_4-y_4dy_2).
\end{split}
\end{equation}
In order to diagonalize the metric, let us perform the following $x^+$ dependent coordinate transformations
\begin{equation}
u=e^{i{m_0q_0x^+\over4}}(y_1+iy_3), \qquad  z=e^{i{m_0q_0x^+\over4}}(y_2+iy_4).
\end{equation}
The final form of the metric is
\begin{equation}\label{iiappmetric}
ds^2 =-4dx^+dx^- - m_0^2\bigg[v^2+{(q_0^2-2)\over8}({\bar u}u+ {\bar
z}z)\bigg](dx^+)^2
+ dx^idx_i + d{\bar u}du+ d{\bar z}dz + dv^2.
\end{equation}
The only nontrivial component of the Ricci tensor is
\begin{equation}
\eqlabel{ricci}
R_{++}={m_0^2q_0^2\over 2}.
\end{equation}
The Penrose limit on the two-form RR field $F_{(2)}= dA_{(1)}$ gives
\begin{equation}
\eqlabel{a1}
e^{\Phi}F_{2}= q_0m_0 dv\wedge dx^+,
\end{equation}
from which we get that the only nontrivial equation of motion in the limit
\begin{equation}
R_{++}={1\over 2}e^{2\Phi}F_{+v}F_{+v}
\end{equation}
is satisfied.

We  rewrite the relevant conserved quantities of this
background anticipating  its field theory interpretation:
\begin{equation}
\eqlabel{conserved}
\begin{split}
H&=i\del_+=i(\del_t+ m_0\del_\phi)= E-m_0J,\\
P^+&={i\over 2}\del_-= {-im_0\over L^2}\del_\phi= {m_0\over L^2}J.
\end{split}
\end{equation}
\subsection{The IIA spectrum}
Let us now consider the IIA GS action  on the
plane wave background given by (\ref{iiappmetric}) and (\ref{a1}).
One of the most important advantages of the Penrose-G\"uven limit we
have performed is that it results in an exactly soluble string
theory. Let us read the bosonic and fermionic worldsheet frequencies.
As usual, we take the light-cone gauge $x^+=\alpha'p^+\tau$. The form of
the metric (\ref{iiappmetric}) directly implies that the bosonic
sector of the system is described by three massless fields with
frequencies $w_n=n$, and five massive (no zero--frequency mode)
fields.  The frequencies for the five massive fields are
\begin{equation}
\eqlabel{freq}
\begin{split}
\omega_n^v &= \sqrt{n^2 + m^2} \\
\omega_n^{u}&= \omega_n^{z}= \sqrt{n^2+{m^2\over8}(q_0^2-2)},
\end{split}
\end{equation}
where $m\equiv m_0\alpha'p^+\approx  J/N^2$.
It is worth mentioning that the three massless fields are also present in
the IIB string spectrum resulting from the Penrose limit of the IR region of the MN and KS
backgrounds \cite{gpss}. They are massless as a result of the Poincare
invariance of the gauge theory. In section \ref{gaugetheory} we will
discuss the interpretation of the string theory Hamiltonian from the
gauge theory point of view, particularly its interpretation as
excitations of dual hadronic states, the so-called annulons.

There is a slightly embarrassing fact. If we restrict the range of the
parameter $q_0\in(0,1)$ in order to have an ALC background in (\ref{gg2}), the $u,z$ zero modes have imaginary
frequencies. Naively this behavior of the frequency would be ruled
pathological, especially for the zero modes since they would seemed to
represent runaway worldsheet modes. However, a closer analysis
\cite{brecher,leomar} reveals that the potential instabilities are
bogus. An analysis of field theory on these backgrounds shows that the
spectrum of energies is generically real even for interacting theories
\cite{leomar}. A perturbative
analysis of supergravity modes in backgrounds with imaginary
worldsheet frequencies also supports the stability of these backgrounds
\cite{brecher}. Overall, the studies carried in similar cases suggest
that the problem might be ultimately an artifact of the light-cone
gauge. Note that similar imaginary frequencies appear even in  the Penrose
limit of such well-behaved systems as some Dp-branes \cite{eric}.

Let us, nevertheless, face up to the fact that some imaginary
frequencies do arise in this limit and try and understand their
origin. As explained in \cite{iiacv,iiab},  $q_0$ is the
parameter that measures the deviation from the $SU(2)^3$ symmetric
metric with $G_2$ holonomy. In other words, $q_0$ governs the
squashing of one of the $S^3$ as well as the fibering of the
space. So, being so intrinsically related to the  angular structure of
the metric, we conclude that most likely  the appearance of imaginary
frequencies reflects our failure to identify the proper $U(1)$ for the
limit.
When  taking the Penrose limit we  tacitly assumed  that the
non-shrinking two-cycle is simply parametrized by $\theta,\phi$. Now,
this is not {\it a priori} obvious since the cycles are really fibered
and the fibering is determined by $q_0$. In
the UV, for example, where the transverse 6-d part of the 10-d metric
describes, at leading order, a standard conifold with $T^{1,1}$ as a
base, the stable two- and three-cycles are defined as $S^2:
\theta=\bar\theta,\, \phi=\bar\phi,\, \psi=0; \,\, S^3: \theta=\phi=0$. Thus,
it is possible that a better identification of the resolved two-cycle
leads to explicitly real frequencies.  However, lacking a clear
criterion for picking the two-cycle, we will not pursue this question
further in this paper.

Concerning the fermions we have that in the light-cone  gauge $\G^+\theta=0$ their equations of motion on the plane wave background at hand
read
\footnote{Here we follow the notations and gamma matrix
conventions of \cite{hyun} and related papers.} (see also appendix
\ref{appendixb})
\be
\begin{split}
(\del_\tau+\del_\s)\theta^1+{m q_0\over4}\gamma^z\theta^2=0,\\
(\del_\tau-\del_\s)\theta^2-{m q_0\over4}\gamma^z\theta^1=0.
\end{split}
\ee
From these equations we find for both $\theta^1,\theta^2$ an equation of the form
\begin{equation}
(\del^2_\tau-\del^2_\s)\theta^a + {m^2 q^2_0\over16}\theta^a =0,
\end{equation}
meaning that all fermions have the same frequencies
\be
\omega^l_n=\sqrt{n^2+{m^2q_0^2\over 16}},\qquad l=1,...8.
\ee
It follows from comparison with the bosonic frequencies  that  the
supersymmetries are not linearly realized on the worldsheet. This is a
universal feature of our string/annulon models  implying a non
trivial zero-point energy \cite{alfra2}.  There are no fermionic zero
modes since $q_0\neq0$, then the  original ${\cal N}=1$ supersymmetry
of the model is not manifest  in the string theory. This is precisely
the same situation encountered for the Penrose limit of the MN
background in \cite{gpss}, where the absence of  linearly
realized supersymmetries was attributed to the failure of finding a
better $U(1)$.

As a consistency check for the above results, it can be easily
verified that in this model $\sum\omega_{Fermion}^2=\sum\omega_{Boson}^2$  order by
order in $n$ implying that our string theory is finite.
\subsection{The zero-point energy}
The zero-point energy plays an important role in the models we
consider. It determines the quantum shift in the energy of the ground
state and the degeneracy of states of the corresponding string
theory.
As mentioned in the previous subsection, our model has a nontrivial zero-point energy which can
formally be written as
\be \label{zeroG2}
E_0(m)={m_0\over 2m}\sum_{n=-\infty}^{\infty}\Biggl[ 3n + \sqrt{n^2 +
m^2}  + 4 \sqrt{n^2+ {m^2\over8}(q_0^2-2)} -
8\sqrt{n^2+{m^2q_0^2\over16}}\Biggr] .
\ee
There are, of course, various ways to regularize the above
expression. In appendix \ref{zeropoint} we discuss two such
regularizations.

As pointed out in \cite{alfra2}, we can evaluate it by regularizing
without renormalizing, a natural procedure for supersymmetric theories.

For small $m$, $E_0(m)$ only gets contributions from the zero
frequencies and so the  imaginary contribution coming from the $u,z$
zero modes appears.
We could have  a real value for the non admitted values $q_0^2\geq2$
for which we would get a negative result for $E_0$.

In the large $m$ limit, the series over $n$ in (\ref{zeroG2})  can
be approximated by an integral over a continuous variable $x\in
(-\infty,\infty)$ and we get $E_0(m)\approx -m_0 m * I$, being $I$ a complex number (for $q_0^2\geq2$, $I$ would instead be real and positive as in all the other models we will consider).

The negativity of the zero point energy is a universal feature \footnote{The model examined in this section is of course a bit problematic in this respect, due to the presence of imaginary frequencies.} of string duals of annulons in supersymmetric or non supersymmetric
confining gauge theories. The non-triviality of $E_0$ here is related
to the fact that only 16 supersymmetries are preserved by the pp-wave
background. For some consideration on the possible meaning of this
result in analogous cases see \cite{alfra,alfra2,alfra3} and also the
general comments we will do at the end of the paper.

\section{The Maldacena-Nastase $2+1$ model and its soft breaking}\label{maldana}
In this section we study the Penrose limit of a supergravity
background that is dual in the low energy regime to ${\cal N}=1$ $SU(N)$
Chern Simons theory. In an attempt to give appropriate credit  to the
authors that have contributed to the construction and understanding of
this gravity dual,  we should mention that the solution was first
constructed by Chamseddine and Volkov in
\cite{Chamseddine:2001hk}. Then,  the brane (ten dimensional)
interpretation was given by Schvellinger and Tran
\cite{Schvellinger:2001ib}.  The gauge theory dual was understood
nicely by Maldacena and Nastase \cite{maldanastase}, and  further
interesting developments can be seen in the work of Gomis
\cite{Gomis:2001xw} and of its softly broken version (bMNa in the
following) obtained in \cite{gae}. Further studies of the  BPS equations and
the ten dimensional Killing spinors can be find in the appendix of
\cite{Nunez:2003cf}.

Let us write some comments on the gauge theory dual. The supersymmetric solution represents D5-branes wrapping a
three-cycle inside a (topologically) $G_2$ holonomy
manifold. This is dual, in the low energy regime,
to a three dimensional gauge theory with two supercharges, the minimal amount in three dimensions.
The twisting leaves us, at low energies,  with a massless bosonic
gauge field and its fermionic superpartner.

From the above comments if follows that the supergravity background we
discuss here and
its dual gauge theory are the lower (3-d) analogue of the MN background.
Recall that the MN background can be viewed as a collection of D5-branes
wrapping a two-cycle. This is one of our main motivations for the study of
the bMNa background.

The supergravity background reads
\bea
\label{MNasol}
ds^2_{str}&=& e^{\Phi } \left[ dx_\mu dx^\mu
+ \alpha' N[ d \rho^2 + R^2(\rho)d\Omega_3^2
+{1 \over 4 } \sum_a (\Sigma^a - A^a)^2 ] \right],
\\
G_3 &=& ie^{\Phi}\alpha'N \left[  -{1\over 4} (\Sigma^1 -A^1)\wedge
(\Sigma^2 - A^2)
\wedge (\Sigma^3-A^3)  + { 1 \over 4}
\sum_a F^a \wedge (\Sigma^a -A^a)\right]+ \nonumber \\
&&+ie^{\Phi}\alpha'N f(\rho)\sigma_1\wedge \sigma_2\wedge \sigma_3
\eea
with a $\rho$-dependent dilaton (whose value at the origin is a
continuum parameter $\Phi_0$), an auxiliary
gauge field $A^a$, its field strength $F^a$ and a
three-form field strength (whose explicit expression depends on $f(\rho)$) given
by
\footnote{Here we implement the symmetry of the IIB equations of motion under the simultaneous changes
$w\rightarrow-\omega, k\rightarrow-k$ on the solution in
\cite{gae}. We prefer
using this ``switched'' version because it has a gauge field $A$ going
to zero in the IR, and this will be useful when  performing  the
Penrose limit (see also \cite{gpss}). The same goal can be reached in
the original solution by a gauge  transformation on $A$.}
\be
\label{Afield}
A^a={1\over 2}(1-w(\rho))\,\sigma^a , \quad  f(\rho)=
-{1\over16}(w^3(\rho)-3w(\rho)+4k), \quad k={1\over2},
\ee
where the last condition follows by imposing regularity at the origin.
The transverse three-sphere is parametrized by Euler angles which we
call $\theta_2, \phi_2, \psi_2$.

One can see by computing the Born Infeld action of a D5 in this
background that the Wess Zumino part of the action contributes
with a factor of the form \beq \frac{1}{16\pi^2}\int d^6x
C_{2}^{RR} F_2 \wedge F_2= (-\frac{1}{16\pi^2}\int d \Omega_3
F_3^{RR} ) \int d^3 x (A\wedge d A + \frac{2}{3} A^3), \eeq that
is, the low energy field theory will contain a supersymmetric
${\cal N}=1$ Chern-Simons term  (apart, of course, from the SYM
action that is suppressed at very low energies). This is a
confining gauge  theory (as the $\rho\to 0$ limit of the metric reflects), that has a
single vacuum. There is a very interesting connection with
supersymmetry breaking, that was  clearly understood in
\cite{maldanastase,Gomis:2001xw}. The main idea is that this field
theory, including massless and massive (KK) fields, has an action
with a Chern-Simons term with level $\kappa=N= \frac{1}{16\pi^2}\int d
\Omega_3  F_3^{RR} $. However, when going to even lower energies
and integrating out the massive fields, the level  of the  Chern-Simons term changes to $\kappa- \frac{N}{2}$.
One can compute the
Witten index and it is zero if $\kappa<\frac{N}{2}$, suggesting that
supersymmetry might be broken in the infinite volume limit. This
is nicely confirmed in the supergravity dual, where, as written
above $\kappa=N$, thus leaving at low energies an $SU(N)$ ${\cal N}=1$
Chern-Simons theory at level $N/2$. If we wrap (in the probe
approximation, which means neglecting backreaction) a small number
$n$ of D5-branes on $S^3$, then  supersymmetry is broken if $n<0$,
that is if we add anti-branes; this reproduces the  breaking
pattern argued above.

In the following, we will specify the functions $w(\rho), R(\rho),
\Phi(\rho)$ near $\rho=0$, that is,  in the dual region to the
IR of the gauge theory. We will introduce a parameter $b$ that takes
values in the interval [0,1).  It turns out that  if $b=\frac{1}{3}$,
we have the supersymmetric solution discussed in
\cite{Chamseddine:2001hk,Schvellinger:2001ib}.
We will consider the case in which we explicitly break the
supersymmetry of the solution, by leaving this parameter $b$ arbitrary
in the interval and we will call this the softly broken MNa solution.

The relevant asymptotics are
\cite{gae}
\begin{eqnarray}
w(\rho) &=& 1 - b \rho^2 + ..., \nonumber\\     R(\rho) &=&\rho -
{(2+9b^2)\over 36}\rho^3 + ...,\\
\Phi(\rho)& =& \Phi_0 + {(2+3b^2)\over8} \rho^2 + ...\nonumber
\end{eqnarray}
Let us note that the one-form field $A^a$ goes
to zero in the IR. In the gauge transformed solution with switched
signs of $w,k$ it goes to a pure gauge. In any case, the
associated field strength goes to zero in the IR.  The range of
allowed values for $b$ is imposed by requiring regularity of the
supergravity solution and its linking with suitable UV
asymptotics. Recall that the value $b=1/3$ corresponds to the
supersymmetric MNa solution.  The other values should correspond
to switching on a gaugino mass term in the dual field theory. Due
to the analogies with the MN case \cite{abcpz}, it is in fact
plausible that $w(\rho)$ plays the role of the dual of the gluino
bilinear.

Let us now shift the flat coordinates as $e^{\Phi_0/2}L^{-1}x_{\mu}\to
x_{\mu}$ where L is an arbitrary constant. This way the tension for
the confining strings of the dual gauge theory reads $T_s =
L^2/(2\pi\alpha')$, while the glueball and KK masses
\footnote{The decoupling of the 3-d YM theory from the KK modes is realized in the limit $e^{\Phi_0}N<<1$. This is beyond the validity of the supergravity approximation
which instead requires $e^{\Phi_0}N>>1$ in order to have small curvatures.} are given by $M_{KK}^2\approx M_{gl}^2\approx L^2/(e^{\Phi_0}N\alpha')$.
Following \cite{gpss}, we will take a Penrose limit of the IR of the supergravity background above, by enforcing the conditions
\be
m_0^2 = {L^2\over e^{\Phi_0}N\alpha'}\quad {\rm{fixed}},\qquad \quad L^2\approx e^{\Phi_0}N\to\infty,
\ee
which send the string tension to infinity while keeping $M_{KK}, M_{gl}$ fixed.

Now we perform the Penrose limit on the ten dimensional background (\ref{MNasol}) along
a null geodesic in the great circle on the transverse $S^3$ defined by $\theta_2=0$ and $\phi_2=\psi_2$ and make the following change of variables
\begin{equation}
x^i \to L x^i, \qquad \rho = {m_0\over L} \,r, \qquad \theta_2 = {2m_0\over L}\,v ,\qquad \phi_+ = {1\over2}(\psi_2+\phi_2).
\end{equation}
This way we get a limit for the IR of the metric in (\ref{MNasol}),
of the form
\begin{eqnarray}
ds^2  &=&  -\,(L^2 +\frac{(2+ 3 b^2)m_0^2 r^2}{8} )dt^2
+\,dr^2 + {r^2\over4}(d\theta_1^2+ d\phi_1^2 + d\psi_1^2 +2\cos\theta_1d \phi_1 d\psi_1) \nonumber\\
&& + (dx_i)^2 + (dv^2+ v^2\,d\phi_2^2)
  + ({L^2\over m_0^2} + \frac{(2+ 3 b^2) r^2}{8}) \,d\phi_+^2 \\
&&  - 2 v^2\,d\phi_2\,d\phi_+
-{b\over2}\,r^2(d\psi_1 + \cos\theta_1 d\phi_1)d\phi_+,\nonumber
\end{eqnarray}
where the new variables $r,v$ have dimension of length.
To reduce the metric in a more diagonal form, let us redefine
\be
\hat \psi_1 = \psi_1-b\, \phi_+, \qquad \quad \hat \phi_2= \phi_2-\phi_+.
\ee
This way we find
\begin{eqnarray}
& &ds^2  =  (L^2 + \frac{(2+ 3 b^2)m_0^2 r^2}{8} )[-dt^2 + {1\over
m_0^2}d\phi_+^2] + dx_i dx^i
+ {dr}^2 +\\
&&{r^2\over4}(d\theta_1^2+ d\phi_1^2 + d{\hat \psi_1}^2
+2\cos\theta_1 d\phi_1 d\hat \psi_1)
+ (dv^2+ v^2\,d\hat\phi_2^2)
   -( {v^2}+
   {{b^2r^2\over4}})\,d\phi_+^2. \nonumber
\end{eqnarray}
Finally we define
\be
x^+=t, \qquad \qquad x^- = {L^2\over 2}(t- {1\over m_0}\,\phi_+),
\ee
and pass to the Cartesian coordinates $d u_1^2 + du_2^2 + du_3^2
+du_4^2= dr^2 +(r^2/4)(d\theta_1^2+ d\phi_1^2 + d{\hat \psi_1}^2 +
2\cos\theta_1 d\phi_1 d\hat \psi_1)$, $\,d v_1^2 + dv_2^2 = dv^2+
v^2\,d\hat\phi_2^2$ .   So, we obtain
\be\label{ppMNr}
ds^2 = -2dx^+dx^- -m_0^2\,[{b^2\over 4}\sum_{j=1}^4u_j^2 + v_1^2 +
v_2^2](dx^+)^2 + dx_idx_i +\sum_{j=1}^4 du_j^2 + dv_1^2+dv_2^2 \ .
\ee
This is very close to the metric obtained in the analogous
four dimensional case \cite{gpss,alfra} from the Penrose limit of the
(b)MN solution and shares the universal features pointed out in the
previous section. The main difference is in the fact that now the
string action on the background (\ref{ppMNr}) will have two massless
scalars ($x_i$) (instead of three as in the 4-d case) and other six
massive ones.  The only difference between the BPS and the broken case
is on the value for the masses of the  scalars $u_j$,  which are
$b$-dependent.  Note that the changing is restricted, just as in the
4-d analogue, to the ``non-universal'' sector of the theory, i.e. the
one  which is not determined by the symmetries of the original
background (the ``universal sector'', which is  parametrized by $x_i,
v_1,v_2$, is in fact $b$-independent).  This is expected, since the
soft-supersymmetry breaking term doesn't change the overall topology
of the metric in the far IR.  As a consequence, the main features of
the field theory annulons will be the same as in the supersymmetric
theory.

In the (b)MN case one gets, as shown in \cite{gpss}, one ``extra''
massless mode (apart from those coming from the three flat space
directions). Here, instead, we have only two massless modes
(coming from the flat directions of space), but the ``extra''
massless mode is absent. This difference seems to appear due to
the fact that here we are twisting the field theory on $S^3$ (to
render it topological) with an $SU(2)$ gauge field, and we need,
by force, to use all the degrees of freedom $A^{1}, A^{2},A^{3}$,
to perform this twist. In the MN case, one could make the twist
with an Abelian field, but, to desingularize the solution, one
needs to turn on the other components of the $SU(2)$ gauge field.
So, it seems that this ``over using'' of the gauge field is
related to the effect of the extra massless mode.

In the Penrose limit on the three-form, only the $\Sigma_1\wedge\Sigma_2\wedge\Sigma_3,\,\, F_3\wedge\Sigma_3$ terms survive.
In particular the components of $G_3$ along the three-sphere on which the D5-branes are wrapped, which give the Chern-Simons
term in the field theory, vanish; this is consistent with the fact that the gauge degrees of freedom, being
uncharged under the internal symmetry we are focusing on, are not seen in the Penrose limit.
All in all we find \footnote{This result is easy to obtain using the expression for the flat 4-d Cartesian
coordinates in terms of the radial coordinate and the Euler angles, see appendix \ref{appendixa}.
}
\be\label{ppH}
 G_3 = -2 i m_0 dx^+\wedge [\,dv_1\wedge dv_2 + {b\over 2}\,( du_1\wedge du_2 - du_3\wedge du_4)\,].
\ee
It is easy to check that the background obtained here satisfies the
supergravity equations
of motion, as ($g^{++}=0$) $R_{++}= {1\over4}(G_{+ij}{G^*}_{+}^{ij})= m_0^2(b^2+2)$.

The string Hamiltonian  and the momentum $P^+$ on the above background are
\bea
\label{MNH}
 H &=& -p_+ = i\partial_+ = E - m_0 (b J_{\psi_1} + J_2 + J_{\psi_2}) \equiv
E-m_0\,J, \nonumber \\
P^+ &=& - {1\over 2} p_- = {i\over 2}\partial_- = {m_0\over L^2}
(b J_{\psi_1} + J_2 + J_{\psi_2}) = {m_0\over L^2} J ,
\eea
where we denote $-i\partial_{\psi_1}$,
$-i\partial_{\phi_2}$ and  $-i\partial_{\psi_2}$ with $J_{\psi_1}$, $J_2$ and $J_{\psi_2}$ respectively.

\subsection{String theory on the bMNa pp-wave}
Studying the string action on the pp-wave background (\ref{ppMNr}), (\ref{ppH}), and choosing the light-cone gauge as usual ($x^+=\alpha'p^+\tau$), produces the following results.

Let us define $m=m_0 \alpha' p^+$. The bosonic sector of the
system is described by two massless fields ($x_i$) with
frequencies $w^i_n=n$, and six massive fields
($u_1,u_2,u_3,u_4,v_1,v_2$) with frequencies \be \omega_n^u =
\sqrt{n^2 + {b^2\over 4}\,m^2},\qquad \quad \omega_n^v = \sqrt{n^2+
m^2}. \ee
It is worth noting that the degeneracy of the latter
frequencies is 4 and 2. Concerning the fermionic sector, we find
eight massive fields whose frequencies are \footnote{To obtain the
result we use a Chevalier decomposition on simultaneous
eigenstates of the operators $i\Gamma_{u_1u_2},\,
i\Gamma_{u_3u_4}, \, i\Gamma_{v_1v_2}$.} (see also appendix B)
\bea
\omega_n^I &=& \sqrt{n^2 + {m^2\over4}} , \qquad  I= 1,2,3,4; \nonumber \\
\omega_n^J &=& \sqrt{n^2 + {m^2\over4}(b+1)^2} , \qquad  J= 5,6;  \\
\omega_n^K &=& \sqrt{n^2 + {m^2\over4}(b-1)^2} , \qquad  K= 7,8. \nonumber
\eea
The sum of the squares of the fermionic frequencies above exactly
matches the sum of the squares of the frequencies of the bosonic
fields order by order in $n$.  Thus the corresponding string theory
is finite both in the susy and the broken case.
As a difference with the analogous 4-d model \cite{gpss,alfra} let
us outline that four fermionic frequencies are $b$-independent, and
so they are not affected by the supersymmetry breaking.


The string zero point energy
$E_0(m)$, evaluated as in the previous section, only gets contributions from the zero frequencies when $m << 1$,  so, for $b\in[0,1)$, we have
\be
E_0(m)\to m_0(b-1) < 0.
\ee
We could have a zero value only for $b=1$, but this is actually excluded.
Anyway, just as for the (b)MN and (b)KS cases \cite{alfra} also for
the supersymmetric
solution $b=1/3$, the zero-point energy is
negative for $m\to 0$. It also stays negative for every value of
$m$. In the large
$m$ limit in fact we get (see appendix D for more details)
\be
E_0(m)\to - {m_0\,m\over4}[2b^2\log b +4\log2 -(b-1)^2\log|b-1| -(b+1)^2\log|b+1|]<0 .
\label{Imn}
\ee
This depends linearly on $m$ and thus $E_0$ takes
larger and larger negative values as $m$ increases.

Let us conclude this section by noticing that $m^2\approx J^2/N^2$ as
for all the other 3-d or 4-d annulon/string models, except the Type IIA 
$G_2$ model of section 3.

\section{D2 and fractional D2-branes}\label{d2fd2}
In this section we analyze the supergravity solution found in \cite{CGLP} 
(CGLP) starting from the presentation done in \cite{herzog}.
The solution in question is  a generalization of the
Type IIA supergravity
solution corresponding to a stack of D2-branes. Besides the fields
corresponding to a stack of
D2-branes the solution includes other fields corresponding to fractional
D2-branes. In this sense this is the 2+1 analog of the
Klebanov-Strassler solution \cite{ks}, where a configuration containing
a stack of D3-branes and a collection of fractional D3-branes was
considered.
The CGLP solutions
are built out of a warped compactification of $2+1$ dimensional
Minkowski space and an asymptotically conical $G_2$ holonomy  manifold
$\Y$.  Hence the metric in the string frame is
\be
ds_{10}^2 = H(r)^{-1/2} dx^\mu dx^\nu \eta_{\mu\nu} +
H(r)^{1/2} ds_\Y^2 \ .
\ee
The variable $r$ is the radius of the asymptotically conical
region of $\Y$.
The two cases considered in \cite{CGLP} correspond to
 $\Y$ being an ${\mathbb R}^{3}$ bundle over a four dimensional Einstein manifold
$M_4$ ($S^4$ or ${\mathbb C}{\mathbb P}^2$).
\be
ds_\Y^2 = \ell^2 \left[ h(r)^2 dr^2 + a(r)^2 (D\mu^i)^2 + b(r)^2 ds_{M_4}^2
\right]\ ,
\label{g2}
\ee
where $\ell$ has dimensions of length and
the $\mu^i$ are coordinates on ${\mathbb R}^{3}$ subject to $\mu^i\mu^i=1$.
The fibration is written in terms of the $SU(2)$ Yang-Mills one-instanton potential $A^i$ where
\be
D\mu^i = d\mu^i+\epsilon_{ijk}A^j \mu^k \ .
\ee
The self dual field strength is denoted by
$J^i =dA^i + \frac{1}{2} \epsilon_{ijk}A^j \wedge A^k$.

The metric (\ref{g2})
is Ricci-flat and has $G_2$ holonomy when the functions
$h$, $a$, and $b$ are given by
\be
h^2 = \left(1-\frac{1}{r^4} \right)^{-1} \; , \; \; \;
a^2 = \frac{1}{4} r^2 \left(1 - \frac{1}{r^4} \right) \; , \; \; \;
b^2 = \frac{1}{2}r^2 \ .
\label{hab}
\ee
The variable $r$ runs from one to infinity.
To highlight the similarity
with the KS solution note that
the parameter
$\ell$ is very similar to the deformation parameter
$\varepsilon$ of the deformed conifold \cite{ks}.
The reader can note that, as in the deformed conifold case, near $r=1$
the highest homology cycle (the four-cycle in this case)  has finite volume
determined by the parameter $\ell$ while the lowest homology cycle
(two-cycle) shrinks to zero size.

As mentioned before, there are various form fields supporting this
metric.  The four-form
RR flux $F_4$ has two pieces, one corresponding to the electric
flux of the ordinary D2-branes aligned in the Minkowski
space-time directions, the other corresponding to magnetic flux
from the ``fractional'' D2-branes.  These fractional D2-branes
are D4-branes wrapped on two-cycles inside $\Y$.  As a
result, they source a flux through a transverse
four-cycle inside $\Y$:
\be
\label{4form}
 F_4 = d^3x \wedge dH^{-1} + m G_4 \ .
\ee
The dilaton is nontrivial, $e^\phi = H^{1/4}$.
A nonzero $m$ forces one to turn on the NSNS three form
flux
\be
\eqlabel{m}
H_3 = m G_3 \ ,
\ee
where $G_3$ is a harmonic three-form inside $\Y$.
The trace of Einstein's equations enforces the condition
on the warp factor
\be
\label{LapH}
\nabla^2 H = -{1 \over 6} m^2 |G_3|^2, \
\ee
where $\nabla^2 $ is the Laplacian with respect to $ds_\Y^2$
and the magnitude $|\cdots|^2$ is also taken with
respect to $ds_\Y^2$. Note that when $M_4 = S^4$, $m=8\pi\alpha'^{3/2}N$, see \cite{herzog}.

The harmonic three-form $G_3$ is
\be
\label{harmonic3}
\ell G_3 = f_1(r) \, dr \wedge X_2 + f_2(r) \, dr \wedge J_2 + f_3(r) \, X_3  ,
\ee
where
\be
X_2 \equiv \frac{1}{2} \epsilon_{ijk} \mu^i D\mu^j \wedge D\mu^k \; , \; \; \;
J_2 \equiv \mu^i J^i \; , \; \; \;
X_3 \equiv D\mu^i \wedge J^i \ .
\ee
The forms obey
$dX_2 = X_3, \,\, dJ_2 = X_3$.
The dual four form $G_4$ is
\be
\label{4formg}
G_4 = f_3 h \, \epsilon_{ijk} \mu^i dr \wedge D\mu^j \wedge J^k +
\frac{f_2 a^2}{h} X_2 \wedge J_2 +
\frac{f_1 b^4}{2ha^2} J_2 \wedge J_2 \ .
\ee
where the functions $f_i$ are
\be
f_1 = ha^2 u_1 \; , \; \; \; f_2 = hb^2 u_2 \; , \; \; \;
f_3 = ab^2 u_3 \ ,
\ee
and the expressions for $u_i$ follows
\begin{eqnarray}
u_1 &=& \frac{1}{r^4} + \frac{P(r)}{r^5(r^4-1)^{1/2}} \ , \nonumber \\
u_2 &=& -\frac{1}{2(r^4-1)} + \frac{P(r)}{r(r^4-1)^{3/2}} \ , \\
u_3 &=& \frac{1}{4r^4(r^4-1)} - \frac{(3r^4-1) P(r)}{4r^5(r^4-1)^{3/2}} \ ,
\nonumber
\end{eqnarray}
where
\be
P(r) = \int_1^r \frac{d\rho}{\sqrt{\rho^4-1}} \ .
\ee
In the
$r\rightarrow 1$ limit, the $u_i$ behave as
\bea
u_1 &\rightarrow & \frac{3}{2} -7(r-1)+{\cal O}((r-1)^2),  \nonumber \\
u_2 &\rightarrow & -\frac{1}{4} + {7\over 10} (r-1) +{\cal O}((r-1)^2),
\nonumber \\
u_3 & \rightarrow & -\frac{1}{4}+ {7\over 5} (r-1) +{\cal O}((r-1)^2).
\eea
It would be convenient to introduce a new radial variable $\tau$ which
is very appropriate for the region $r\approx 1$, according to
\be
\eqlabel{tau}
d\tau \sim dr /2\sqrt{r-1}.
\ee
Next, we calculate the warp factor which is given by
\be
H(r) = \frac{m^2}{2 l^6} \int_r^\infty \rho
\left( 2u_2(\rho) u_3(\rho) - 3u_3(\rho) \right) d\rho \ .
\label{Hformula}
\ee
The integration constant has been chosen such that $H(r) \sim \frac{Q}{r^5}$
in the limit $r\rightarrow \infty$.
In the other limit, $r=1$,
\be
H(r)\approx \frac{m^2}{l^6} (a_0-{7\over16}(r-1))\equiv h_0 - h_1\tau^2
\label{h_1}
\ee
where $a_0 \approx  0.10693$.

The fact that $H(r)$ becomes a constant at small
$r-1$ was the original reason motivating the belief
that the gauge theory dual is confining. Note the absence of a linear
term in $\tau$ which would have prevented us from taking the Penrose
limit near the tip of the $G_2$ space.

Let us make some brief comments that help establish the structure
of the  gauge theory dual. The following comments are  based on
the  two previous investigations related to this background,
\cite{herzog} that deals with IR aspects of the gauge theory and
\cite{Loewy:2002hu} dealing with UV  aspects. Let us remind some
background work. It has been suggested, from field theory
considerations \cite{ds,Hanany:1997hr}, that in ${\cal
N}=1$ SYM the tension of the string ending on $q$ external quarks
is proportional to $\sin(\pi q/N)$.  This dependence was rederived
using AdS/CFT methods in \cite{klebanovherzog}. \cite{herzog} used
as main tool the confining strings of the gauge theory and showed,
by probing the geometry with few D4-branes (wrapping $S^3$ inside
$M_4 = S^4$), that the tension of this probe (that is the tension
of the confining string in the gauge dual) displays all the
properties of  a confining string in $SU(N)$ gauge theory. Indeed,
it vanishes when the number of quarks is  $q=0$ or $q=N$, is
symmetric under the change $q\to N-q$, and has the desired
convexity behavior. So, it was argued that the gauge theory is
confining, with ${\cal N} =1$ supersymmetry and with a Chern-Simons term due to the four-form field $F_4$ in $M_4$.
The  level
of the CS term is classically $\kappa=N/2$ and, by integrating out
massive fields, the coefficient is changed to $\kappa=0$. We will be
dealing with adjoint hadrons in this theory.  The situation is
much less clear when the manifold is $M_4={\mathbb C}{\mathbb P}^2$.

On the UV side, that is, when viewed as an M-theory solution, Loewy and
Oz \cite{Loewy:2002hu} have  computed the energy momentum tensor two-point function by the usual AdS/CFT methods and argued that, in that
regime the field theory looks like ${\cal N}=1$ $SU(N)$ gauge theory
coupled to a real scalar with a  quartic superpotential. Since this will
not be the regime our hadrons will explore, we refer the  reader to the
nice paper \cite{Loewy:2002hu} for further details.

\subsection{The Penrose-G\"uven limit}
In this section we perform the Penrose-G\"uven limit of the CGLP
solution. We will specialize to the case $M_4=S^4$. One of the
ingredients that is necessary  to take the limit of  the solution
is an explicit form of the metric on $S^4$ and the corresponding
one-instanton gauge field (see appendix \ref{appendixc}). Here we
find it convenient to use the following parametrization for the
sphere \be ds_{S^4}^2 = d\psi^2 +{1\over 4} \sin^2 \psi
\big[\s_1^2 +\s_2^2 +\s_3^2 \big]. \ee In these coordinates the
one-instanton field can be written in terms of the spin connection
$(A^i=-\omega_{0i}-{1\over 2} \epsilon_{ijk} \omega_{jk})$
\cite{egh} and is given explicitly as \be A^i=-{1\over
2}(1-\cos\psi)\,\,\s_i,\qquad {\rm i=1,2,3}. \ee A standard
parametrization for the $\mu^i$'s is \be
\mu^1=\sin\alpha\sin\beta, \qquad \mu^2=\sin\alpha\cos\beta,
\qquad \mu^3=\cos\alpha. \ee With this notation the $G_2$ metric
takes the explicit form
\bea
\frac{1}{\ell}ds^2_\Y&=&h^2 dr^2 +
b^2(r)\bigg[d\psi^2 +{1\over 4} \sin^2 \psi(\s_1^2
+\s_2^2 +\s_3^2)\bigg]+\nonumber \\
&+& a^2(r)(d\a -\cos\b \, A^1+\sin\b\, A^2)^2+ \nonumber \\
&+& a^2(r)\sin^2\a\, (d\b + \sin\b\cot\a\,\, A^1 +\cos\b\,\cot\a\,\, A^2
-A^3)^2.
\eea
Parametrizing the Euler angles by $(\theta, \phi_1, \phi_2)$ we obtain
\be
\s_1^2+\s_2^2 +\s_3^2 = d\theta^2 +d\phi_1^2 +d\phi_2^2+2\cos\theta
d\phi_1 d\phi_2.
\ee
We find it convenient to further introduce
\be
\phi^{\pm}={1\over 2}(\phi_1\pm \phi_2).
\ee
We  would like to explore a sector of the IR of the dual gauge theory by
performing a Penrose limit on the dual supergravity background. We do not have an
explicit radius/energy relation, but we are confident that going
to $r\to1$ means going to IR in the dual 3-d model (see the analogous
discussions for the 4-d case previously examined).
The null geodesic we want to zoom in is determined by the following
conditions on the coordinates: $\psi={\pi\over 2},\quad \te=0$. In the
radial position we will focus on $r=1$ which in the $\tau$ coordinate
(\ref{tau}) corresponds to $\tau=0$. Following Penrose's prescription we expand up to
quadratic terms near this null geodesic by introducing a suitable parameter $L$.  As usual
we first shift the $2+1$ coordinates as $x^{\mu}\to h_0^{1/4}L x^{\mu}$
and send $L\to\infty$ while keeping
\be
m_0^2 \equiv {2L^2\over{\ell^2h_0^{1/2}}}\quad \rm{fixed}.
\label{mzero}
\ee
Note that $h_0= a_0m^2/\ell^6\approx N^2$ and so in the limit $L^2\approx N$.
We then re-shift $x^i\to x^i/L$ and take
\be
\psi={\pi\over 2}-{m_0\over L}v_3 ,\quad \te= {2m_0\over L}y, \quad
\tau={m_0\over{\sqrt2} L}\rho.
\ee
Expanding the whole metric near this null geodesic gives
\bea
ds^2&=& L^2[-dt^2 + {d\phi_+^2\over m_0^2}] -m_0^2\rho^2{h_1\over 4h_0} dt^2 + dx_idx_i + dw_jdw_j  +
 2(w_2dw_1-w_1dw_2)d\phi_+\nonumber \\
&&  + dv_3^2 + dy^2+y^2d\phi_-^2-[v_3^2 -(2-{h_1\over 4h_0})(w_1^2+w_2^2)-(1-{h_1\over 4h_0})w_3^2 + y^2]d\phi_+^2,\nonumber
\eea
where $i=1,2;\,\, j=1,2,3;\,\, w_j=\rho\mu_j$. Now, in order to diagonalize the previous expression, let us introduce the complex variables
\be
u=e^{-i\phi_+}(w_1+iw_2), \qquad v=ye^{i\phi_-},
\ee
and the light-cone coordinates
\be
t=x^+, \qquad x^-= {L^2\over2}(t-{\phi_+\over m_0}).
\ee
This way we get the final expression (we rename $w_3\to z$)
\bea
ds^2&=& -4dx^+dx^- - m_0^2[v_3^2+v{\bar v} +({h_1\over 2h_0}-1)(u{\bar u} + z^2)]dx^+dx^+ +\nonumber \\
&& dx_idx_i + dud{\bar u} + dzdz + dvd{\bar v} + dv_3^2.
\label{pmetf}
\eea
Thus, after taking the light-cone gauge, this metric will describe
two massless worldsheet bosons arising from the
spatial worldvolume directions; three bosonic fields with unit mass
arising from three of the four coordinates of the original $S^4$; and
three equal mass bosonic fields arising from the original radial
direction and the shrinking two-cycle.
However, this conclusion is premature since the background includes a
nontrivial $B$-field that will slightly modify it and prevents us from
extending the above counting beyond the $n=0$ sector.

Let us now turn to the rest of the background. The dilaton goes to a
constant $e^{\Phi_0} = h_0^{1/4}$.

In the expression for the harmonic three-form (\ref{harmonic3}), the first
term vanishes in the limit. Since $X_{(3)}=dJ_{(2)}$, the last two term
combine in the limit to give
\be
H_3 = -{m\over 8\ell}{m_0\over{\sqrt{2}L}}[dw^j\wedge J^j],
\ee
%
and so, after some algebra and using (\ref{h_1}), (\ref{mzero}), we find
\be
H_3=-{m_0\over2}\sqrt{{h_1\over 14h_0}}dx^+\wedge[2dv_3\wedge dz + i(du\wedge d{\bar v}-d{\bar u}\wedge dv)].
\label{3-forma}
\ee
The four-form field strength has two terms (\ref{4form}).  The
first one describes the regular D2-branes and vanishes in the limit.
This is similar to the vanishing of the regular D3-branes in the limit of the
Klebanov-Strassler solution.
The second structure in the four-form is proportional to $G_4$
(\ref{4formg}). In the limit all its terms contribute.
What we find is that $F_4$ goes as (see also eqns. $(26,27)$ in \cite{herzog}; also note that $dJ^k$ terms are subleading)
\be
F_4 = {3\over{16}}m\,J_{2}\wedge J_{2}+ m {m_0^2\over{32L^2}} \epsilon_{jkl}d[w_j\wedge dw_k\wedge J^l],
\ee
and so we finally get
\be
F_4 = -{m_0h_0^{-{1\over4}}\over2}\sqrt{{h_1\over 14h_0}}dx^+\wedge[idv_3\wedge(-6dv\wedge d{\bar v}+ du\wedge d{\bar u}) + dz\wedge(du\wedge d{\bar v} + d{\bar u}\wedge dv)],
\label{4-forma}
\ee
the $dv\wedge d{\bar v}$ term originating from the $J_{2}\wedge J_{2}$ piece.
It is now easy to check that the only non-trivial equation for the whole pp-wave background at hand
\be
R_{++}= {1\over12}e^{2\Phi_0}F_{+abc}{\bar F}_{+abc} + {1\over4}H_{+ab}{\bar H}_{+ab}
\ee
is satisfied.

The conserved quantities $H$ and $P^+$ read as usual
\bea
\label{d2fd2h}
H &=&i\partial_+ = i\partial_t + im_0\partial_{\phi^+}
= i\partial_t + im_0(\partial_{\phi^1}+ \partial_{\phi^2})
= E -m_0J,\nonumber \\
P^+ &=&{i\over2}\partial_{-}= -i{m_0\over L^2}\partial_{\phi^+}
= -i{m_0\over L^2}(\partial_{\phi^1}+ \partial_{\phi^2})= {m_0\over L^2}J.
\eea

\subsection{The spectrum}
Let us now study the IIA GS string on the above background. There are two
massless
bosonic fields corresponding to the two spatial directions $x_i, i=1,2$.
Due to the presence of the NSNS field, the bosonic equations of motion
for the six massive fields are not diagonal. Nevertheless they can be
easily solved finding that the corresponding $3+3$ frequencies read
(here $m=m_0\alpha'p^+$, do not confuse it with the  $m$ introduced with
the fluxes in (\ref{m}))
\bea
{\omega^p_n}^2 &=& n^2 + {m^2h_1\over{4h_0}}+ \sqrt{m^4\left(1-{h_1\over{4h_0}}\right)^2 + m^2 n^2{h_1\over 14h_0}}, \qquad p=3,4,5; \nonumber \\
{\omega^q_n}^2 &=& n^2 +{m^2h_1\over{4h_0}}- \sqrt{m^4\left(1-{h_1\over{4h_0}}\right)^2 + m^2 n^2{h_1\over 14h_0}}, \qquad q=6,7,8.
\eea
Thus the six massive fields are effectively assembled in two groups.
The sum of the squares of the bosonic frequencies reads $\sum_{B} \omega_n^2= 8n^2 +(3/2)m^2(h_1/h_0)$.

The fermionic case is as usual less trivial. The equations of motion read 
(see also appendix B)
\bea
(\partial_{\tau} + \partial_{\sigma})\theta^1 &=& {\alpha'p^+\over{(4) 2!}}H_{+ij}\Gamma^{ij}\theta^1 - {\alpha'p^+\over{(4) 3!}}e^{\Phi}F_{+ijk}\Gamma^{ijk}\theta^2\equiv B\theta^1 + A\theta^2, \nonumber \\
(\partial_{\tau} - \partial_{\sigma})\theta^2 &=& -{\alpha'p^+\over{(4) 2!}}H_{+ij}\Gamma^{ij}\theta^2 - {\alpha'p^+\over{(4) 3!}}e^{\Phi}F_{+ijk}\Gamma^{ijk}\theta^1\equiv -B\theta^2 + A\theta^1.
\eea
Now, let us expand the fields as $\theta^a=\sum_n \theta_n^a(\tau)e^{in\sigma},\quad a=1,2$ , substitute in the equations of motion and take the derivative respect to $\tau$. Then introduce a complex field $\epsilon_n(\tau) = \theta^1_n(\tau) + i\theta^2_n(\tau)$. Let us also note that the $A, B$ terms commute. All in all we arrive to an equation of the form
\be
\partial^2_{\tau}\epsilon_n = [A^2 + (B-in)^2]\epsilon_n.
\ee
Let us now find solutions for the fermionic equations using the
following Chevalier decomposition:
\be
i\Gamma^{u_1\,v_2}\xi^{\pm\, .\, .}=\pm\,\xi^{\pm\, . \, .},\quad i\Gamma^{x\,w_3}\xi^{.\, \pm\, .}=\pm\,\xi^{.\,\pm\, .},\quad i\Gamma^{v_1\,u_2}\xi^{.\, .\,\pm}=\pm\,\xi^{. \, . \,\pm}.
\ee
This way we obtain the following frequencies
\bea
{\omega^p_n}^2 &=& n^2 + {13m^2\over16}{h_1\over 7h_0} +{m\over4}\sqrt{2h_1\over 7h_0}n , \qquad p=1,2,3; \nonumber \\
{\omega^q_n}^2 &=& n^2 + {13m^2\over16}{h_1\over 7h_0} -{m\over4}\sqrt{2h_1\over7 h_0}n , \qquad q=4,5,6; \nonumber \\
{\omega^7_n}^2 &=& n^2 + {45m^2\over16}{h_1\over 7h_0} +{3m\over4}\sqrt{2h_1\over7 h_0}n ; \nonumber \\
{\omega^8_n}^2 &=& n^2 + {45m^2\over16}{h_1\over 7h_0} -{3m\over4}\sqrt{2h_1\over 7h_0}n .
\eea
It follows that the sum of the squares of the fermionic
frequencies is $\sum_{F}\omega_n^2= 8n^2 + (3/2)m^2(h_1/h_0)$
which exactly matches with the analogous bosonic sum, ensuring finiteness to our model.

The zero-point energy is non trivial also in this case, sharing
the general features of the model examined in section 4.
Finally, note that $m^2\approx J^2/N^2$.
\section{Comments on gauge theory interpretation}\label{gaugetheory}
In this section we translate the string theory results in terms of
the dual gauge theory. Let us first address the ground state
$H|\Omega \rangle =0$, where $H$ is  the string theory Hamiltonian in
terms of gauge theory quantities. According to (\ref{conserved})
for the resolved conifold with RR two-form, (\ref{MNH}) for the
softly broken MNa and (\ref{d2fd2h}) for the background describing
a collection of regular  and fractional  D2-branes, the ground
state has energy $E-E_0=m_0J$. Moreover, since in the light-cone
quantization $P^+$ is a finite nonzero constant, we see that in
the limit $(L\to \infty)$ the global charge  $J$ has to be very
large. Thus, the gauge theory ground state has a very large mass
$E$ proportional to its global charge $J$. We call this state, as
in \cite{gpss}, annulon.

The string Hamiltonian, properly interpreted, describes excitations of
the annulons.  The bosonic string Hamiltonian in all the cases
examined in the paper can be written explicitly following standard
manipulations. Here we provide the needed notation to understand its
form.  We define number operators

\bea
\N_R&=&\sum_{n=1}^\infty n \big( a^{i\dagger}_{n} a_{n}^i \big) \
,\ \ \ \ \N_L=\sum_{n=1}^\infty n \big( \tilde a^{i\dagger}_{n}
\tilde a^i_{n}
\big)\ ,\nonumber \\
N_R &=&\sum_{n=1}^\infty n \big( a^{s\dagger}_{n} a^s_{n} \big) \ , \
\ \ \ N_L=\sum_{n=1}^\infty n \big( \tilde a^{s\dagger}_{n}
\tilde a^s_{n} \big)\ ,
\eea
and sub--Hamiltonians
\bea
H_0&=&  \omega_0^s \big(a^{s\dagger}_{0} a_{0}^s
\big) \ , \nonumber \\
H_R &=& \sum_{n=1}^\infty \omega_n^s \big(a^{s\dagger}_{n} a_{n}^s
\big),\quad \, H_L=\sum_{n=1}^\infty \omega_n^s \big(
\tilde{a}^{s\dagger}_{n}
\tilde{a}_{n}^s \big) \ .
\eea
The subindex $i=1,2,3$ (or $i=1,2$) refers to the three (two) flat
directions in the plane wave (spatial directions in the gauge theory),
while the index $s=4,5,6,7,8$, ($s=3,4,5,6,7,8$) runs over the
internal directions. There is implied summation over the indices $i$
and $s$. The full bosonic light-cone Hamiltonian is
\bea
\label{TheHamiltonian}
H=-P^- &=& H_{\parallel} + H_{\perp} \\ &=& \left[{P_{i}^2 \over 2P^+
} + {1\over 2 \a' P^+}\left(\N_R+
\N_L\right)\right]
+ \left[{1\over 2 \a' P^+}\left(H_0+ H_R+ H_L  \right)\right]
\nonumber\ .
\eea
The string theory Hamiltonian is thus constructed of a contribution from the
momentum and massless stringy excitations in the directions
(index $i$), $H_{\parallel}$, and a contribution from the
massive ``zero'' modes and excitations of the  directions
(index $s$), $H_{\perp}$.

Let us now interpret the string theory Hamiltonian (\ref{TheHamiltonian}) in terms
of field theory quantities. The first term in  $H_{\parallel}$
describes the nonrelativistic motion of the annulon. This can be seen
by using the dictionary between string theory and gauge theory:
$P^+=m_0J/(2\pi\a'T_s)$ and $P_i^2 (2\pi\a'T_s)={\cal P}_i^2$. The
first entry can be read explicitly from  the definition of the scaling
(\ref{stringscale}) and the particular form of $P^+$ in the
backgrounds we considered: (\ref{conserved}) for the
resolved conifold with RR two-form, (\ref{MNH}) for the softly broken
MNa and (\ref{d2fd2h}) for the background describing a collection of
regular  and fractional  D2-branes. The relation between the momenta
is simply a result of the relative warping between ten- and
four-dimensional quantities. Thus, on the gauge theory side  the first term in
$H_{\parallel}$ can be written as ${\cal P}_i^2/(2m_0 J)$ which
therefore represents the nonrelativistic free motion of the annulons
in the directions $i$. The second term in $H_{\parallel}$ describes
typical stringy excitations.

It is very satisfying to see that the annulons of IIA share precisely
the same sector $H_{\parallel}$ with the annulons of IIB. We thus
verify that this is a truly generic feature of these theories dictated
by the symmetries of the backgrounds. Moreover, part of the universal
sector of the annulons of IIB and IIA is in the oscillations in the
$v$ directions, as we will see momentarily.

\subsection{The ``Universal Sector''}
We call universal the sector of the Hamiltonian (\ref{TheHamiltonian})
that is completely determined by the symmetries of the background. As
we have mentioned, $H_{\parallel}$ is completely determined by the
symmetries of the problem. We see that the  directions $i$ come
directly from the spatial directions of the
worldvolume  of the stack of branes where the gauge theory lives. Some of the terms in $H_{\perp}$ are
also determined by symmetries and we will described them in this section.

The are various reasons why making a precise identification of the
string theory spectra with the field theory data is difficult. One
reason is the lack of the  state/operator
correspondence in nonconformal theories. Another, perhaps more
important reason is that since we are dealing with string states which
are not protected by supersymmetry the
``anomalous energies'' \footnote{Recall that we work on the Poincare
patch and therefore the eigenvalues of the operator dual to time is
not the dimension of an operator but the energy of a given state.}  can be very large.
Even in cases where the gauge theory is relatively well known
as in the  (b)KS and (b)MN  studied in   \cite{gpss,alfra}, the
matching is very challenging.

Let us recall, once again, the structure of the field theory we want
to describe.
We are concerned with the study of supergravity backgrounds dual to
confining theories in three or four dimensions, which contain  pure (S)YM,
in some cases  with Chern-Simons terms, coupled to many other fields.
A crucial property of the supergravity backgrounds is that they are non-singular near
the origin of the space transverse to the flat coordinates.  This fact
allows us to study objects localized in the IR of the field theory, with a
large internal global charge $J$, the annulons.  Since the gauge fields are
not charged under the global $U(1)$ symmetry, the annulons must be created by the other
fields present in these theories.  Among these, the lightest scalars
turn out to be the essential ingredient. These lightest scalars determine part of the
``universal sector'' of the sting theory, i.e. the sector  constrained by
the symmetries of the supergravity background.

These scalars have a simple origin.  The models we have at hand
are all believed to come from some configurations of Dp-branes.
Let ``p'' stand for the dimensionality of the brane before any
wrapping. For example p=5 for the (b)KS, (b)MN and (b)MNa, p=6 for
the 4-d IIA case and p=4 for the CGLP one.  Then, before wrapping,
the world-volume theories contain 9--p scalar fields, which are the
ones we are interested in.  Upon wrapping the branes on cycles, of
vanishing size for the fractional case, and eventual twisting of
the world-volume theory, all these scalars get  masses
(generically, they get the same mass).  Note that in the process
they can combine among themselves, as in the KS model, but the
number of scalar degrees of freedom is unaltered.  So, in all our
theories we expect that the lightest charged objects, which can
account for (part of) the string theory universal sector, are
simply the 9--p scalars describing the motion of the original
branes in ten dimensional flat spacetime.  Note that they are all
in the adjoint representation of the gauge group.

The identification goes on as follows.  The backgrounds we
consider are non-singular because they have a non vanishing cycle
at the origin.  They are the non-singular geometries created in
the large $N$ limit by $N$ Dp-branes wrapped on cycles.  Due to
the wrapping, in the large $N$ limit, the flat transverse space
develops some finite volume cycle which supports the flux.   The
corresponding flux is the magnetic (8--p)-flux generated by the
Dp-branes. For example, in the (b)KS or (b)MN models we have a
three-form flux supporting the non-vanishing three-cycle of the
deformed  conifold, in the 4-d IIA theory a two-form flux through
the two-cycle of the resolved conifold, and so on. So, in general
we have an (8--p)-cycle. Since the $U(1)$ direction required to
take the Penrose limit is taken from this cycle we conclude that
this (8--p)-cycle determines the rest of the universal sector.

As we have just explained,  this cycle is the compactification
of the original flat space transverse to the branes,
and so it is ``made of'' our scalar fields, which thus naturally
transform under the internal global $U(1)$ charge  associated to the equator of the cycle.
To be precise, the cycle accounts for 7--p combinations of the scalars as coordinates on it, plus another one giving
its radius. The scalars corresponding to cycle modes have naturally  unit mass in the appropriate units.
It is natural to associate the $U(1)$ directions used to take the
limit with  one of the scalar fields that has unit charge. Thus,
the annulon can be viewed
as the hadron created by many (large $U(1)$ charge equal $J$) such scalars.
Since  the annulon is a charged massive object, there is
also a state of charge $-J$, which cannot be seen on the
string side, this one being focused on the $+J$ states.
But its presence reflects itself in the duality in the fact
that a second scalar field, which would be the one to create the
$-J$ state, is typically unstable once inserted in the annulon
ground state, so that we loose its degree of freedom in
the string description.
This leaves us with 7--p scalar degrees of freedom, to be matched
with the remaining 7--p directions of the cycle orthogonal
to the geodesic.
The latter are always present in our plane-waves as unit mass
(in $m_0$ units) world-sheet bosons: they are the $v$ modes.

The universal sector is completed by the flat coordinates $x_i$. Other
bosonic modes,  forming the ``non-universal sector'', are very hard to identify even when the dual
field theory is better known \cite{gpss,alfra}. For this reason we
will not purse the nonuniversal sector.
Some fermionic fields dual to the world-sheet zero mode spinors can be
possibly identified within the (would be superpartners  in the
broken supersymmetry cases) superpartners of the above scalar fields.

Thus, even if we do not know the details of the field theories dual to
our supergravity backgrounds, we have enough
informations to identify the universal sector.

\subsubsection{Annulons in 4-d and 3-d theories}
Let us consider explicitly the identification of some string theory
states with gauge theory states for the  string theory quantities computed in
Section 3,4 and 5.

We start by considering the 4-d IIA case.
In taking the Penrose limit we picked a geodesic direction and  one transverse coordinate,
the $v$ one in the metric (\ref{iiappmetric}).
On the dual field theory side, we have three massive scalar
fields, coming from the ones parametrizing the motion of the D6-brane
on flat space-time.
Let us first see how the fields on the stack of D6-branes transform under the global symmetry
group $SO(1,6) \times SO(3)$ (we first consider flat D6-branes):
\beq
A_\mu= (7,1), \qquad \Phi^a= (1,3), \qquad \Psi= (8,2).
\eeq
When the D6-branes wrap the three-cycle 
(with normal bundle $SU(2)$) the isometry group
is $SO(1,3)\times SU(2)_\Sigma \times SU(2)$  
and the bosonic fields transform as (no twisting yet)
\beq
A_\mu=(4,1,1) + (1,3,1), \qquad \Phi^a=(1,1,3).
\eeq
After the twisting takes place, the first 
will be the massless gluon and the second and third massive
scalars.
For the fermions, we have
\beq
\Psi=(2,1,2)+ (2,3,2) + cc
\eeq
giving place, after the twisting, to massive fermions and the gluino.

Here we shall consider the way in which the fields transform under
a $U(1)$ inside $SU(2)$. Let us focus, for reasons explained below,
on the scalars coming from $\Phi^a$. These three scalars are in a
triplet of $SU(2)\sim SO(3)$, so they have charges $+1,\,0,\,-1$
with respect to the $U(1) \subset SU(2)$.

From the discussion above, we see that the $J=+1$ scalar,  call it
$A_1$, creates the ground state annulon, dual to the geodesic
direction. Namely, the string theory ground state is dual to the
field theory hadron which is the lowest energy state obtained by
acting with the operator ${\rm Tr}\,A_1^J$ on the field theory
vacuum: ${\rm Tr}\,A_1^J|0\rangle$, where $|0\rangle$ is the field
theory vacuum. Note that to first approximation in the large $J$
limit, the state ${\rm Tr}\, A_1^J|0\rangle$ has mass equal to
$Jm_0$.  A second scalar degree of freedom, the $J=0$ one $A_0$,
is dual to the $v$ direction, i.e. its insertion in the string of
$A_1$'s above is dual to the state obtained by the action of the
$v$ zero mode creation operator on the vacuum. Its Hamiltonian in
fact reads $H=E_0+m_0$. Finally, the third scalar cannot be seen
in string theory, because it is dual to the $-J$ state.

The identification of the $u,\,\bar{u},\,z,\,\bar{z}\,$
string zero-modes would require a more accurate knowledge of the dual
field theory. Even in the case of conformal theories, as the dual to
$AdS_5\times T^{1,1}$ discussed in \cite{susyenhanced,Gomis:2002km}, the field
theoretic description of these modes involves antichiral fields. \\

Let us examine more closely the 3-d models. The universal sector of the (b)MNa theory has,
apart from the two $x_i$ components, the two $v_i$ modes which have, as
expected, mass equal to $m_0$.
The theory comes from the wrapping of five-branes in IIB on a three-cycle.
In analogy with the previous decomposition, we start by recalling
how the fields on the brane transform under the isometries
$SO(1,2)\times SU(2)_\Sigma \times SU(2)_L\times SU(2)_R$,
once the brane wraps the three sphere. Before implementing the twisting,
one has
\beq
A_\mu= (3,1,1,1) + (1, 3,1,1),\qquad \Phi^a=(1,1,2,2),
\label{descomposicionbosones}
\eeq
and for the fermions
\beq
\Psi=(2,2,2,1)+ (2,2,1,2) .
\eeq
Similarly to the previous case, the twisting mixes $SU(2)_\Sigma$ with $SU(2)_L$ into
$SU(2)_D$ in a way that preserves minimal supersymmetry. We will have
fields  that transform under this $SU(2)_D$
(massive fields)  and some other fields  that are inert under it
(massless fields). The  angular momentum of the massive fields,
according to eq. (\ref{MNH}), is given by $J=  b J_{\psi_1} + J_2 +
J_{\psi_2}$, being $J_{\psi_1} $ the charge under the $U(1)$ inside
$SU(2)_\Sigma$,  $J_2$ the charge of the  $U(1)$ inside  $SU(2)_L$ and
$J_{\psi_2}$ of the one in $SU(2)_R$.  We will have massive bosons
coming from the decomposition of the gauge field $(1,3,1,1)$ in
(\ref{descomposicionbosones}); these have $J=\pm b,0$. The massive
excitations coming from the scalars can be named according to their
charge under $U(1)_{\psi_1} \times U(1)_{L} \times U(1)_{R} $  as
\beq
A_{+,+}= (0,+,+), \;\; A_{-,-}= (0,-,-), \;\; A_{-,+}= (0,-,+), \;\; A_{+,-}= (0,+,-).
\eeq
Thus, we see that
$A_{++}$, $A_{--}$, $A_{-+}$ and $A_{+-}$ have $J$ charge $1,\,-1,\,0,\,0$ respectively.

We are led to the conclusion that the operator ${\rm Tr}\,
(A_{++})^{J}$ acting on the field theory vacuum is dual to the
vacuum of the string theory, i.e. it is the annulon in the ground
state.  Its Hamiltonian has the appropriate value $H=0$.  The
modes $v,\, \bar{v}$ ($=v_1\pm i v_2$) are instead identified with
(the insertion in the string of $A_{++}$ of) $A_{+-}$ and
$A_{-+}$. The states obtained acting with the $v,\bar v$ zero-mode
creation operators have in fact Hamiltonian $H=E_0+m_0$ and are
conjugate to each other. Finally, as usual $A_{--}$ is unstable,
when inserted in the string of
$A_{++}$, against decay to the other scalars \cite{gpss}.  \\

Having established the pattern of fields explicitly in the previous two
examples, we will be brief with the field theory dual to the CGLP
background. There are three unit-mass string
zero-modes coming from the  four-sphere.  The theory is a fractional D2
model, where the fractional branes are wrapped D4's, giving five
transverse scalars transforming in $SO(5)$ to begin with.  With respect
to the $U(1)$ corresponding to $\Phi_+$ (which is a $U(1)$ in a
$SU(2)$), there should be two uncharged scalars and three in a triplet,
so one $J=1$ scalar giving the string vacuum, three $J=0$ scalars giving
the three modes of the universal sector $v_3,\,v,\,\bar{v}$, and the
decoupled $J=-1$ one.

\subsection{Zero point energy and Hagedorn behavior}
The complete Hamiltonian is obtained by summing to the bosonic one
(\ref{TheHamiltonian}) the fermionic contributions. We do not
write explicitly the whole Hamiltonian, but concentrate for the
moment on the total zero point energy $E_0(m)$. All the models we
investigate trough this paper have non-trivial zero point energy,
which is generically negative (actually it is complex in the
resolved conifold case) and diverging in the large $m$ limit
\cite{alfra2} (we refer the reader to appendix D for a detailed discussion on this result).

Let us here go to an interpretation of the string zero point energy for the models at hand in view of the string-annulon correspondence. 

Let us consider for example the string
ground state and the gauge theory annulon corresponding to it. The
string Hamiltonian for this state is $H=E_0(m)$ and so the prediction
for the corresponding hadronic energy (mass) is $E = m_0J +
E_0(m)$. Now, if we think the annulon constituents to have bare mass
$m_0$ and unit $J$-charge, the classical prediction for these should be
simply $E_c = m_0J$. So, in some sense string theory is giving us a
prediction on the bare masses and the quantum corrections to
them. Let us see it in more detail.

The crucial ingredient to understand this is the following. The
parameter $m$ is related to field theory quantities by $m= aJ/g_sN=
1/\lambda'$ (with $a$ a model-dependent numerical constant) being
$\lambda'$ (in the standard BMN notations where $\lambda'=\lambda/J$, with $\lambda=g_sN$ ) the effective coupling for
the sector we are analyzing \footnote{Note that this relation holds in
all the confining (4-d and 3-d) models apart from the 4-d IIA in the
section 3 of this paper where $m\approx J/g_s^{2/3}N^2$.}. 
Now, the
perturbative field theory regime is reached when $m>>1$. In this limit
generically $E_0(m)\to -m_0 m A^2$ , and so
\be
E\approx m_0J - m_0A^2m = m_0J( 1- {aA^2\over g_sN})
\ee
which shows that in the large $N$ limit we recover the classical
expectation $E = m_0J$, but from the point of view of the theory in
the Penrose limit, which measures order one fluctuation around the
$E=m_0J$ configuration, the second term cannot be ignored. Read in
terms of the effective coupling $\lambda'$, we have $E\approx m_0 (J -
A^2/\lambda')$ and so the zero-point term can be read as a
nonperturbative (in $\lambda'$) field theory contribution to the annulon masses.
Since $m_0$ cannot be computed in the perturbative field theory, the
presence of the above corrections does not change much the situation.  So,
in principle, we could just accept the new value of $m_0$ as a stringy
prediction and go on to calculate, for example, the corrections to the
mass of the annulons with insertions of field corresponding to string
oscillators, along the line of BMN.

In appendix D we will trace a difference between our way of evaluating $E_0$ and the Casimir renormalization prescription often found in literature. Let us note here that the latter would predict that the classical result $E_c=m_0J$ is only modified by exponentially suppressed terms in the large $m$ limit. So the two
methods give in fact different predictions for the nonperturbative
contributions to the annulon mass, a fact that in principle offers to our choice the privilege of falsifiability.

In the strong effective coupling limit, $m\to0$, $E_0\to -k^2m_0$ ,
being $k$ a model-dependent constant. In this case $E\approx m_0J -
k^2m_0 = m_0J (1- k^2/J)$ and so the prediction is that for $J>>1$ we
still get $E=m_0J$.

Finally, let us spend a few words on the thermodynamics of the string
models we are considering in this paper. Thanks to their exact
solubility, we can evaluate the string partition function at finite
temperature and search for the typical Hagedorn behavior for hadronic
matter. Both the Casimir prescription and the
non-renormalizing one preserve modular invariance (see \cite{alfra2})
and so this is sensible indeed.

The calculation
is standard \footnote{See, for example, \cite{leovam}.}, and as usual the
Hagedorn temperature is given by the zero point energy evaluated with
anti-periodic fermions.  It is given implicitly by \cite{alfra,alfra2}
($T_H=\beta_H^{-1}$)
\be
\beta_H= -2\sqrt{2}\pi\alpha' E_{tot}\left({m_0\beta_H\over2\sqrt{2}\pi};0,1/2\right),
\label{hagmn}
\ee
where $E_{tot}(M;0,1/2)$ is the zero-point energy $E_0(M)$ evaluated
with periodic bosons and antiperiodic fermionic fields. Note that due to
the strategy we are adopting, i.e. calculating $E_0$ without
renormalizing it, the above result differs from the one in
\cite{pandovaman} for a finite term in $E_{tot}$.

\subsection{Experimental signature: Annulon trajectories}\label{experiment}
\begin{figure}
\begin{center}
\epsfig{file=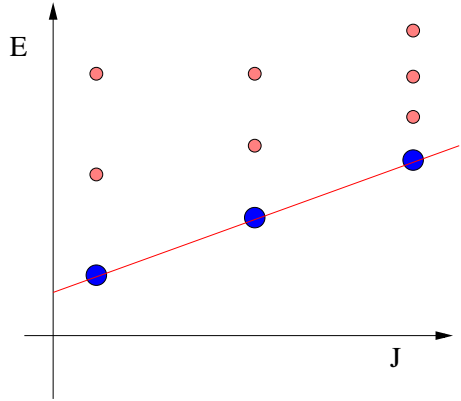,width=6cm}
\caption{ The annulon trajectory obtained as a plot of the lowest mass
annulon for a given charge.\label{antr}}
\end{center}
\end{figure}

Regge trajectories are one of the most clear experimental signatures of
the spectrum of  particles
involved in strong interactions. Since we are considering hadronic bound
states, we are motivated to think about a
``smoking gun'' type of experimental signature for  the
annulons. The quantum numbers that describe our bound states are the
energy $E$ and the global $U(1)$ charge $J$. Thus, it makes sense to
introduce a modified
Chew-Frautschi plot where on the y-axis we will have the energy
$E=\sqrt{t}$ of the annulon and on the  x-axis we  have its charge $J$.
By inspecting the Hamiltonian (\ref{TheHamiltonian}) and
recalling the quantum numbers for the ground state we see that in
general we have
\be
\label{trajectory}
E=m_0 J + \alpha_0 + {\rm excitations}.
\ee
As can be seen from Fig. \ref{antr}, for a given  charge $J$ we have a whole
tower of states that represent the excitations of the ground state
annulon. More remarkably, in some similarity to the Regge trajectories,
{\it the  plot containing the lowest energy annulon for a given charge
$J$ is  a straight line}.
Some remarks are in order. The expression (\ref{trajectory}) is
the same for all annulons discussed in this and in previous papers
and is therefore  universal. There is an implicit assumption that
$J$ and $E$ are very large quantum numbers so that the
semiclassical analysis can be trusted. The factor $\alpha_0$ is
determined by the zero point energy discussed in the previous
section. As the excited state energy, it is a function of $J$ in
the ratio $J/\lambda$. Since both $J$ and $\lambda$ are
parametrically large, it gives a subleading correction (in $J$) to
the linear relation (\ref{trajectory}). Note that for higher
values of the charge $J$, the distance between excited states in
the same tower decreases.

These annulon trajectories are different from the Regge trajectories
where one obtains a straight line in a plot of spin versus
mass square, $E^2$. Interestingly, the appearance of $E$ rather than
$E^2$ took place in a recent analysis of Regge trajectories in the
context of the gauge/gravity correspondence \cite{regge}.

\section{Non-supersymmetric deformations of the resolved conifold}\label{nonsusyiia}
In section 3, we analyzed the IIA background
corresponding to a resolved conifold metric
with RR two-form flux over the blown up $S^2$.
This is dual to ${\cal N}=1$ SYM and is obtained
by reduction of an eleven dimensional
background which contains a $G_2$ holonomy metric
in the $D_7$ family \cite{iiacv,iiab}.

In this section, we  look  for non-supersymmetric
deformations of the above background.
In particular, we would like to find
seven dimensional metrics
whose asymptotics are similar to the supersymmetric ones.
The ansatz is exactly the same as Eq.(\ref{gg2}), namely
\begin{align}
ds^2 & = d r^2 +
a(r)^2 \left[ (\Sigma_1 + g(r) \sigma_1)^2
+ (\Sigma_2 + g(r) \sigma_2)^2 \right]
\nonumber \\
& + c(r)^2 (\Sigma_3 + g_3(r) \sigma_3)^2
+ b(r)^2 \left[ \sigma_1^2 + \sigma_2^2 \right]
+ f(r)^2 \sigma_3^2\,.
\label{ansatzCGLP}\end{align}
The results are as follows.
We found a family of regular non-supersymmetric solutions
with  $g(r) \equiv 0$
but otherwise  essentially the same small $r$ behavior
as the supersymmetric ones.

As in the supersymmetric case, upon reduction to Type IIA, the dilaton is still
finite everywhere.
However, the large $r$ asymptotic properties
differ dramatically from the supersymmetric case.
In particular, there are {\it two} stabilized one-cycles instead
of one. The IIA metric in string frame reads
\bea
\label{metriiiaNSUSY}
&&e^{-2\Phi/3} ds_{10}^2 = dx_{1,3}^2 + \ell^2\bigl[ dr^2 + a(r)^2 (d\bar{\theta}^2 +
\sin^2\bar{\theta}d\bar{\phi}^2)
+ b(r)^2 (d\theta^2 +
\sin^2\theta d\phi^2) +\nonumber \\
&&+\frac{f^2 c^2}{f(r)^2 + (1 + g_3(r))^2
c(r)^2}
(d\psi_2 -
\cos\theta d\phi + \cos\bar{\theta} d\bar{\phi})^2 \bigr]\nonumber,
\eea
and the radius of the circle parametrized by $\psi_2$ has
a finite limit as $r \to \infty$.
The matter fields are given as before by
\bea
\label{matteriiaNSUSY}
&&A_1= \ell N\bigg[\frac{f(r)^2 - (1
- g_3(r)^2) c^2}{f(r)^2 + (1 +
g_3(r))^2 c(r)^2} (-d\psi_2 +
\cos\theta d\phi - \cos\bar{\theta} d\bar{\phi}) + \nonumber \\
&& \qquad \qquad +\cos\theta d\phi +
\cos\bar{\theta} d\bar{\phi}\bigg],\nonumber \\
&&4N^2e^{4\Phi/3}= f(r)^2 + c(r)^2 (1 + g_3(r))^2.
\eea
Ricci flatness implies
that the functions $\{a(r),b(r),c(r),f(r),g(r),g_3(r)\}$
satisfy a coupled system of six second order
nonlinear differential equations supplemented
by a constraint. The precise form of these equations can be found in
appendix \ref{appendixd}.
From consistency with the equations of motion and the
requirement that the solution be regular
at $r=0$, one can derive the following boundary conditions
\begin{align}
b & = R_0
- \left( \frac{q_0^2-2}{16 R_0} \right)
r^2 + {\cal O}(r^4)\,,
\nonumber
\\
f & = q_0 R_0 + \frac{q_0^3}{16 R_0} r^2
+ {\cal O}(r^4) \,,
\nonumber
\\
a & = \frac{1}{2} r
+ \left(
\frac{q_0^2 - 4 R_0^2}{192 R_0^2} + \frac{c_0}{2}
\right) r^3
+ {\cal O}(r^5)\,,
\nonumber
\\
c & = - \frac{1}{2} r
+ c_0 r^3 + {\cal O}(r^5)\,,
\nonumber
\\
g & = \frac{q_0}{2} + \alpha
+ \left[
\frac{(q_0 + 2 \alpha) \left( (q_0^2 + \beta)^2
+ q_0^2 (\alpha^2 + \alpha q_0 - 1) \right)}{24 q_0^2 R_0^2}
\right] r^2 + {\cal O}(r^4)\,,
\nonumber
\\
g_3 & = \left( \frac{q_0^2}{2}-1+\beta \right) +
\left[
\frac{ (4 \alpha^2 + 4 \alpha q_0)(q_0^2 + \beta)
+ q_0^2 ( 2 q_0^2 - 2 + 3\beta)}{24 R_0^2} \right] r^2
+ {\cal O}(r^4)\,.
\label{boundaryCGLP}\end{align}
The supersymmetric solutions correspond to
\begin{align}
c_0 = -\frac{5 q_0^2 - 8}{288 R_0^2}\,,
\quad \alpha = 0\,, \quad \beta = 0\,.
\end{align}
In particular, these boundary conditions imply that
$g(r) = -\frac{a f}{2 b c},\, g_3(r) = 2 g(r)^2 - 1$ for any value of $r$.

The only non-singular non-supersymmetric solutions we were able to
find correspond to setting $\alpha = -\frac{q_0}{2}$.
Note that this is not a small deviation
from the supersymmetric case. In fact,
by virtue of the equations of motion,
$\alpha =  -\frac{q_0}{2}$ implies that $g(r) \equiv 0$.
The value of $c_0$ is still
the same as in the supersymmetric case.
Thus, besides the usual parameters $q_0$ and $R_0$,
these solutions depend on the value of $\beta$.
Numerical analysis shows that there is actually
only a finite range of values
of $h_0 \equiv \frac{q_0^2}{2}-1+\beta$ such that
the solution is regular and finite everywhere
(\,see Fig. \ref{range}\,).
If $h_0$ lies outside this range,
the solution develops a singularity
at a finite value of $r$.
Note also that, since
the equations of motion are symmetric under
$g_3(r) \to - g_3(r)$, which is true
only if $g(r) \equiv 0$,
the range of allowed values for
$h_0$ is symmetric under $h_0 \to - h_0$.

These solutions are markedly different from
the supersymmetric ones.
First of all, both $f(r)$ and $c(r)$ approach
a constant value in the limit $r \to \infty$.
As a consequence, the Type IIA metric is not asymptotically
conical as in the supersymmetric case. In fact,
the fibered $S^1$ parametrized
by $\psi_2$ has actually a finite radius
at infinity.
As $r \to \infty$, a generic solution
has the following behavior
\begin{align}
a(r) & = \frac{r + r_\infty}{\sqrt{3}}
+ \frac{3 \sqrt{3} \left(\,
q_2^2 ( 1 + 2 g_\infty^2 ) + 2
\,\right) R_2^2}{8 (\,r+r_\infty)}
- \frac{c_3 + q_2 f_3}
{2 \sqrt{3} q_2 R_2 (\,r+r_\infty)^2} + ...
\nonumber \\
b(r) & = \frac{r + r_\infty}{\sqrt{3}}
+ \frac{3 \sqrt{3}
\left(\,
q_2^2 ( 2 + g_\infty^2 ) + 1
\,\right) R_2^2
 }{8 (\,r + r_\infty)}
- \frac{c_3 + q_2 f_3}
{2 \sqrt{3} q_2 R_2 (\,r+r_\infty)^2} + ...
\nonumber \\
c(r) & = q_2 R_2
- \frac{9 q_2^3 R_2^3 (1 + g_\infty^2)}
{4 (\,r+r_\infty)^2}
+ \frac{c_3}{(\,r+r_\infty)^3}
+ ...
\nonumber \\
f(r) & = R_2 - \frac{9 R_2^3}{4 (\,r+r_\infty)^2}
+ \frac{f_3}{(\,r+r_\infty)^3} + ...
\nonumber \\
g_3(r) & = g_\infty
- \frac{9 R_2^2 g_\infty}{2 (\,r+r_\infty)^2}
+ \frac{ h_\infty }{(\,r+r_\infty )^3}
+ ...
\label{asymptinftyFINAL}\end{align}
where the seven parameters involved,
$\{ q_2, R_2, c_3, f_3, g_\infty, h_\infty,
r_\infty \}$, will depend on the three IR parameters $\{ q_0, R_0, h_0\}$.
%
The dilaton is still finite everywhere,
just like in the supersymmetric case.

The RR two-form flux
through the non-collapsing two-sphere at infinity
parametrized by $\theta$ and $\phi$
is not equal to $1$ as in the supersymmetric case.
This is because the first term in the RR gauge field
(\ref{matteriiaNSUSY}) proportional to $\sigma_3 - \Sigma_3$, namely
$A(r) \equiv \frac{f^2 - (1-g_3)^2 c^2}{f^2 + (1+g_3)^2 c^2}$,
does not vanish for $r \to \infty$ as in the supersymmetric case.
Actually, in the limit $r \to \infty$, it
always goes to a {\it negative} constant,
which in general is not even an integer.
However, the flux through the two-sphere defined by
$\psi_2 = 0, \theta = \bar \theta, \phi = \bar \phi$
is always equal to $2N$, independently of
the limit of $A(r)$. This holds for supersymmetric solutions too.
Note that this sphere does not collapse either,
since its area is proportional to $b(r)^2 + a(r)^2$.

To create a more detailed and intuitive picture  we show a generic
solution corresponding to
$\{ R_0=1, q_0=\frac{1}{2}, h_0=-\frac{1}{2} \}$ in figures
 \ref{cippa1}-\ref{cippa6}.
Fig. \ref{cippa6} shows that the term
$A(r) \equiv \frac{f^2 - (1-g_3)^2 c^2}{f^2 + (1+g_3)^2 c^2}$
is not asymptotic to an integer constant.
Setting $h_0 = -0.395$ instead, we find
a solution such that $\lim_{r \to \infty} A(r) = -2$
(\,Fig. \ref{cippa7}\,).
The asymptotic values of $e^\Phi$ and $c(r)$ are
different in the two cases
(\,Figs. \ref{cippa8},\, \ref{cippa9})\,,
whereas the profiles of $a(r),b(r)$ and $f(r)$ are
virtually the same.
In Figs. \ref{cippaAB}-\ref{cippaRR}, we compare
the supersymmetric solution for $\{ R_0=1, q_0=\frac{1}{2} \}$
and the non-supersymmetric one corresponding
to $\{ R_0=1, q_0=\frac{1}{2}, h_0=-0.395 \}$.

Interestingly, the Penrose limit,  with the particular
geodesic of section 3, of
these non-supersymmetric solutions
is exactly the same as the Penrose limit of the supersymmetric ones with the same value of $q_0$.
This enhancement of supersymmetry  for the Penrose limit of certain supergravity
backgrounds was first noticed in \cite{susyenhanced,Gomis:2002km}
for  the case of
$AdS_5\times T^{1,1}$ which has the same Penrose limit as $AdS_5\times
S^5$. Basically,  the Penrose limit of a general solution whose
behavior close to $r=0$ is given by
Eqs.(\ref{boundaryCGLP})
is exactly the same as the supersymmetric one.
The metric reads
\begin{align}
ds^2&=-4dx^+dx^- - m_0^2\bigg[v^2+ \frac{(2 q_0^2-4(1+g_0^2))}{16}
y^ay_a \bigg](dx^+)^2+dx^idx_i+dy^ady_a+dv^2
\nonumber \\
&+m_0 g_0 dx^+(y_1dy_3-y_3dy_1+y_2dy_4-y_4dy_2)\,,
\label{metricG}\end{align}
where $g_0 \equiv \frac{q_0}{2} + \alpha$.
For the family of non-supersymmetric
solutions that we found, one has $g_0 \equiv 0$.
In general, with the following coordinate transformation
$$
u=e^{i \frac{m_0g_0x^+}{2}}(y_1+iy_3),
\qquad  z=e^{i \frac{m_0g_0x^+}{2}}(y_2+iy_4),
$$
we retrieve Eq.(\ref{iiappmetric}). This property of supersymmetry
enhancement in the Penrose limit  gives further  evidence that the
nonsupersymmetric deformation is different from the other soft susy
breaking
achieved by means of a gaugino bilinear. This kind of supersymmetry
breaking seems to be similar to the breaking between the ${\cal N}=4$
SYM corresponding to string
theory on $AdS_5\times S^5$ and the ${\cal N}=1$ SCFT dual to $AdS_5 \times T^{1,1}$.

In summary, for given values of $q_0$ and $R_0$,
there is a one-parameter family of non-supersymmetric
solutions whose flux through the non-collapsing
two-sphere defined by $\psi_2=0, \theta = \bar \theta,
\phi = \bar \phi$ is equal to $2N$, like in the
supersymmetric case.
The asymptotic behavior of these solutions differs
from the supersymmetric case in that the IIA metric
is not conical for large $r$ but has a finite radius
circle instead.

\section{Conclusions}\label{conc}
Let us summarize  our main results  and comment on possible extensions
or the work presented here.

First we should emphasize,  once again, that this paper deals with
supergravity duals to confining gauge theories. For
the supergravity approximation  to be well defined we need the curvature of
the metric to be small in string units.  Small curvature implies that
the field theory dual is not simply that of the `pure' gauge
theory. Instead, the dual necessarily includes  adjoint massive fields
(KK modes). These inconvenient `impurities' plague all the known gravity duals to confining
gauge theories.

As mentioned  in the introduction, this work focuses on the ``hadrons''
composed out of a large number of the  adjoint massive fields referred
above. These hadrons are shown to be ubiquitous in duals to confining
theories, and
some common features are investigated in this paper.

To address some of the dynamical features of these hadrons we used a
Penrose limit that was motivated in \cite{gpss}. Basically, we used a
null geodesic localized in the region of small values of the radial
coordinate. By taking this Penrose limit we found a parallel plane wave
associated to a sector of the confining dual.
Then, quantizing the Type II superstring on the pp-wave
we analyzed several features common to all these confining
backgrounds.  Among these features we discussed: a mass formula for the
ground state and its excitations, the  composition of the hadrons
in terms of KK modes, the existence of `universal' sectors, zero point
energies, and a experimental signature for the annulon trajectories.

It is worth highlighting that we considered backgrounds dual to
minimally supersymmetric field theories, that is, with four supercharges in
four dimensions and two supercharges in three dimensions. We also
considered backgrounds where supersymmetry is explicitly  broken.
The breaking procedure is `soft' in
the 3-d model of section 4, in
the sense that it is based on the de-tuning of  one parameter; this is not the case for the 4-d model of section 7.
The general structure of our hadrons does not seem to be too sensitive
to the breaking. In fact, it also coincides with the same general type
studied before. The persistence of this structure for backgrounds in IIB,
IIA, dual to confining gauge theories in  four and three dimensions,
encourages us to view  this fact as a sign of the universality of the
hadrons.

There are, of course, some differences in the several cases that we
considered.
In the case of duals to confining three dimensional gauge theories (Type
IIB or IIA cases), everything seems to work nicely. In the M-theory or
IIA  set up dual to four dimensional  ${\cal N}=1$ SYM + KK
modes there are issues with the lower modes. These problems do not seem
to indicate an instability in the free strings case, but the introduction
of interactions could render the plane wave geometry unstable. This is
obviously one of the features that deserve further study.


Other than the universality of annulons we have established
the existence of a Hagedorn density of states for
the strings quantized in these backgrounds. This is the expected
behavior for hadronic matter. On the more technical side of this
question, we have included a discussion (hopefully fairly clear) on the
issue of regularizations of the zero point energies of the
backgrounds under study; this can be find in appendix D.

In section 7, we have numerically constructed {\it new solutions} that are
non supersymmetric versions of $G_2$ holonomy manifolds in M-theory (or
Type IIA with fluxes on the resolved conifold). These solutions are ALC,
thus having a consistent IIA interpretation as branes wrapping cycles in a
non-supersymmetric fashion. It would be of interest to study properties
of this solutions in relation to their argued field theory dual.
The stability of these new solutions is
also a point to be studied.

What are the future directions of research that this investigation opens?
In view  of their  universality, these hadrons deserve more study. Indeed, the present paper and
its predecessors seem to be the only quantitative prediction for the KK
modes. The scattering properties, a detailed study of the annulon
trajectories discussed in section \ref{experiment}, seem to
be natural directions to concentrate further efforts.

\section*{Acknowledgments}
We would like to acknowledge conversations with Daniel Freedman, Gary Gibbons, 
Sean Hartnoll, Ruben Portugues, Cobi Sonnenschein, Yasutaka Takanishi and Diana Vaman.
G. Bertoldi  is supported by the Foundation
BLANCEFLOR Boncompagni-Ludovisi, n\'ee Bildt.
F. Bigazzi is partially supported by INFN.
C. N\`u\~nez is supported by a Pappalardo Fellowship and in part by
funds provided by the U.S. Department of Energy (D.O.E) under cooperative research agreement
DF-FC02-94ER40818.
L. Pando Zayas is supported in part by DoE.

\appendix
\section{Parametrization of $\mathbb{R}^4$}\label{appendixa}

The left-invariant one forms of $SU(2)$ are defined as
\be
\s_1+i\s_2= e^{-i\psi}(d\theta +i\sin\theta d\phi),
\qquad \s_3 = d\psi +\cos\theta d\phi.
\ee
Note that they satisfy  $d\s_1=-
\s_2\wedge \s_3$.

The change of coordinates on $\mathbb{R}^4$ used in the main text is
\begin{eqnarray}
y_1 = r\sin{\theta\over2}\sin{{\psi-\phi}\over2}, \quad y_2 = r\sin{\theta\over2}\cos{{\psi-\phi}\over2},\nonumber \\
y_3 = r\cos{\theta\over2}\cos{{\psi+\phi}\over2}, \quad y_4 = r\cos{\theta\over2}\sin{{\psi+\phi}\over2}.
\end{eqnarray}
Note that with this definitions the metric in $\mathbb{R}^4$ is
\begin{equation}
ds^2=dr^2 +{r^2\over 4}(\sigma_1^2+\sigma_2^2+\sigma_3^2)=dy^a dy_a,
\end{equation}
with $a=1,2,3,4$.

\section{Fermionic equations of motion}\label{appendixb}
In the generic Type IIA case the equations of motion of the world-sheet
fermions read (see for example \cite{parvizi} and related papers)
\bea
(\partial_{\tau} + \partial_{\sigma})\theta^1 &=&{\alpha'p^+\over{4}}
e^{\Phi}F_{+i}\Gamma^{i}\theta^2
+{\alpha'p^+\over{8}}H_{+ij}\Gamma^{ij}\theta^1
- {\alpha'p^+\over{24}}e^{\Phi}F_{+ijk}\Gamma^{ijk}\theta^2,\nonumber \\
(\partial_{\tau} - \partial_{\sigma})\theta^2 &=&-{\alpha'p^+\over{4}}
e^{\Phi}F_{+i}\Gamma^{i}\theta^1-{\alpha'p^+\over{8}}H_{+ij}\Gamma^{ij}\theta^2
- {\alpha'p^+\over{24}}e^{\Phi}F_{+ijk}\Gamma^{ijk}\theta^1.
\eea
The relevant IIB equations are \cite{gpss}
\bea
(\partial_{\tau} + \partial_{\sigma})\theta^1 &=&-{\alpha'p^+\over{8}}
e^{\Phi}F_{+ij}\Gamma^{ij}\theta^2 -{\alpha'p^+\over{8}}H_{+ij}\Gamma^{ij}\theta^1,\nonumber \\
(\partial_{\tau} - \partial_{\sigma})\theta^2 &=&-{\alpha'p^+\over{8}}
e^{\Phi}F_{+ij}\Gamma^{ij}\theta^1+{\alpha'p^+\over{8}}H_{+ij}\Gamma^{ij}\theta^2.
\eea

\section{The BPST Instanton}\label{appendixc}

In this appendix we present the explicit form of the one-instanton field
that we use in the main text. We present the coordinate change of
variables that takes the explicit solution of \cite{egh} into the one
used in the main text. The metric on $S^4$ and the correspondent
one-instanton are \cite{egh}:
\bea
ds^2&=& {1\over (1+ r^2/a^2)^2}\bigg[dr^2 + {1\over 4}r^2 (\s_1^2
+\s_2^2 +\s_3^2)\bigg], \nonumber \\
A^1&=&-\s_1\, {(r/a)^2\over 1+r^2/a^2},\qquad {\rm cyclic}.
\eea
Taking
\be
\sin \psi = {2 r/a\over 1+r^2/a^2},
\ee
we find
\bea
ds^2&=&{a^2\over 4}\bigg[d\psi^2+ {1\over 4}\sin^2\psi(\s_1^2
+\s_2^2 +\s_3^2)\bigg], \nonumber \\
A^1&=&-{1\over 2}(1-\cos\psi)\,\,\s_1,\qquad {\rm cyclic}.
\eea
These are the expressions used in the main text. The radius of $S^4$ or
alternatively the size of the instanton is set to $a=2$.

The corresponding field strength is given by
$J^i =dA^i + \frac{1}{2} \epsilon_{ijk}A^j \wedge A^k$
\be
J^1=-{1\over 2}\sin\psi d\psi \wedge \s_1-{1\over 4}\sin^2\psi\,\,
\s_2\wedge \s_3, \qquad {\rm cyclic}.
\ee

\section{Zero point energy}\label{zeropoint}
In this appendix we will briefly discuss two of the procedures most commonly encountered in the literature to deal with the zero point energy. We attempt to
describe their similarities and differences. We hope this appendix
clarifies why our preference falls on the one used in the main text.

The Penrose limits we took in this paper always resulted in plane-waves preserving only 16 supersymmetries. These are the so-called ``kinematical''
supercharges, which give no linearly realized supersymmetry on the
worldsheet \cite{cve}. This is evident by examining the various string
spectra we found: worldsheet bosons and fermions have generically
different worldsheet masses. Thus the zero point energy $E_0(m)$ is not trivial.
In particular two aspects of it are relevant: its negative sign and its actual behavior as a function of $m$. As we will show, the first is {\it independent} on
the way in which $E_0$ is evaluated, while the latter is strongly
dependent on it.
 

The general way in which we wrote $E_0$ is in terms of an expression like \footnote{In the regular plus fractional D2 model, as in the KS case and its soft breaking \cite{gpss,alfra}, the frequencies have  in general a more complicate structure. We focus here on the simpler case, just to discuss our evaluation philosophy avoiding other technical complications.}
\be
S(m)={1\over2}\sum_{n=-\infty}^\infty [\sum_{i=1}^8\sqrt{n^2+ b_i^2m^2}-\sum_{j=1}^8\sqrt{n^2+ f_i^2m^2}],
\label{zerotot}
\ee
where $b_i$ and $f_i$ parameterize the bosonic and fermionic masses respectively. The supergravity equations of motion crucially imply $\sum_i b_i^2=\sum_j f_j^2$. Under this condition it is easy to show that the above series is convergent, has definite sign (negative in all the models we examined, apart from the by now problematic IIA resolved conifold one) and is linear in $m^2$ in the large $m$ limit. We will return on these points in a moment. 
There is in fact a formal subtlety here we want to alert the reader about. Since the single bosonic and fermionic contributions of the whole expression above are divergent, we cannot trivially say that the expression in (\ref{zerotot}) is equal to
\be
S_B(m)-S_F(m)={1\over2}\sum_{n=-\infty}^\infty \sum_{i=1}^8\sqrt{n^2+ b_i^2m^2}-{1\over2}\sum_{n=-\infty}^\infty \sum_{j=1}^8\sqrt{n^2+ f_j^2m^2}.
\label{zerotot2}
\ee
If we write $E_0$ in these terms, we have to find a regularization prescription to carefully deal with the divergent terms
\be
{1\over2}\sum_{n=-\infty}^\infty \sqrt{n^2+m^2}.
\label{zero}
\ee
One way is to regularize these is by considering 
\be
\sum\limits_{n\in\mathbb{Z}}\sqrt{n^2+m^2} \longrightarrow
\sum\limits_{n\in\mathbb{Z}}(n^2+m^2)^{-s}.
\ee
Let us thus evaluate this expression
\bea
\label{reg}
&&{1\over 2}\sum\limits_{n\in \mathbb{Z}}(n^2+m^2)^{-s}=\nonumber\\
&&\qquad={1\over 2}\sum\limits_{n\in \mathbb{Z}}{1\over
  \Gamma(s)}\int\limits_0^\infty dt
t^{s-1}e^{-t(m^2+n^2)}\nonumber \\
&& \qquad={1\over 2}\sum\limits_{p\in \mathbb{Z}}{\pi^{1/2}\over
  \Gamma(s)}\int\limits_0^\infty dt t^{s-3/2}e^{-t m^2 -\pi^2
  p^2/t}\nonumber\\
&& \qquad={1\over 2}{\pi^{1/2}\over \Gamma(s)} \left(
2\sum\limits_{p=1}^\infty \int\limits_0^\infty dt t^{s-3/2}e^{-t m^2 -\pi^2
  p^2/t}+ \int\limits_0^\infty dt t^{s-3/2}e^{-t m^2}\right)\\
&& \qquad ={1\over 2}{\pi^{1/2}\over \Gamma(s)} 
2   \sum\limits_{p=1}^\infty 2 \left({\pi^2 p^2\over m^2}\right)^{s/2-1/4}K_{s-1/2}(2\pi\;p\;m)
+{1\over 2}{\pi^{-s}\over \Gamma(s)}   \int\limits_0^\infty dt \;t^{s-3/2}e^{-t m^2} \nonumber \\
&&\qquad =2{\pi^s(m^2)^{-s/2+1/4}\over\Gamma(s)} 
  \sum\limits_{p=1}^\infty {1\over p^{1/2-s}}K_{s-1/2}(2\pi\;p\;m) + {1\over
    2}{\pi^{1/2}\over \Gamma(s)} (m^2)^{-s+1/2}\; \Gamma(s-1/2).\nonumber
\eea
In step 1 we  used  \cite{GradshteynRyzhik} 
\be
\int\limits_0^\infty dt \; t^{\nu-1}e^{-\mu t} =\mu^{-\nu}\Gamma(\nu),
\ee
with $\mu=m^2+n^2$ and $\nu=s$. Note that the $\Gamma$ function
provides a natural analytical continuation for $\nu=-1/2$. 
Step 2 uses the Poisson resummation formula:
\be
\sum\limits_{n\in Z}e^{-t\;n^2}=\left({\pi\over t}\right)^{1/2}\sum\limits_{p\in
  Z}e^{-\pi^2 p^2/t}.
\ee
In step 4 we used \cite{GradshteynRyzhik} 
\be
\int\limits_0^\infty dt t^{\nu-1} e^{-t\gamma-\beta/t}= 2\left({\beta\over
  \gamma}\right)^{\nu/2}K_{\nu}(2\sqrt{\gamma\beta}). 
\ee
The above expression (\ref{reg}) is divergent in the limit $s\to
-1/2$. 

A prescription commonly found in the literature
consists in defining (\ref{zero}) by assigning  it
the finite value corresponding to ignoring the second term in the last
expression in (\ref{reg}). Basically this defines a ``renormalized'' (Casimir energy like) version of (\ref{zero}) to be the term containing the sum of Bessel functions. If we apply this prescription to each term in (\ref{zerotot2}) and then sum up, we find that the whole zero point energy is given by a sum over Bessel functions only. 

In the main body of the paper we followed instead a different ``regularization'' prescription as suggested in \cite{alfra2}. We did not take the
$s\to -1/2$ limit for each individual degree of freedom. Instead we
first added the contribution of bosons and fermions as in (\ref{zerotot}) and only at the end we sent $s\to -1/2$. These two
prescriptions give two slightly different results. 
 

Renormalizing individual degrees of freedom is justified and necessary in the bosonic string cases in order to deal with infinities as we do in ordinary bosonic field theories. A deeper
reason for this is provided by the fact that in a bosonic theory the
zero point or Casimir energy is an observable and thus we renormalize it to obtain a
finite result. In the supersymmetric cases the natural objects to look at are the boson-fermion constituents and the observable quantity is the one obtained by summing the contributions of all degrees of freedom as in (\ref{zerotot}). In a sense this choice amounts on choosing the same regulator (say, the same cutoff) for all the degrees of freedom, and this is physically sensible. This way no renormalization prescription is required: we only need a calculation tool to evaluate a convergent quantity.

In our case, we can thus decide to evaluate the expression in (\ref{zerotot}) by using the $s$-regularization as above, i.e. taking the $s\to-1/2$ limit of $S(m,s)$. 
Due to the condition $\sum_i b_i^2=\sum_j f_j^2$, it follows that  the leading
divergence coming from $\Gamma(-1+\epsilon)$ with $\epsilon \to 0$ in (\ref{reg}) always cancels in $S$. The final result is
\bea
S(m)&=& \sum_{i=1}^8 [2{(b_im)\over \pi^{1/2}\Gamma(-1/2)} 
  \sum\limits_{p=1}^\infty {1\over p}K_{-1}(2\pi\;p\;b_im)]+ \nonumber\\
 &-& \sum_{j=1}^8 [2{(f_j m)\over \pi^{1/2}\Gamma(-1/2)} 
  \sum\limits_{p=1}^\infty {1\over p}K_{-1}(2\pi\;p\;f_jm)]+\nonumber \\
&-&{m^2\over2}[\sum_{i=1}^8 b_i^2\log b_i^2- \sum_{j=1}^8 f_j^2\log f_j^2],
\label{zerototus}
\eea  
the last line \footnote{The origin of this finite correction can be traced to the
fact that 
\be
\lim\limits_{\epsilon\to 0}\,\,(m^2)^{1-\epsilon}\;\Gamma(-1+\epsilon) =
m^2\left({-1\over \epsilon}-1 +\gamma\right) + m^2 \log m^2 + {\cal
  O}(\epsilon).
\ee} tracing the difference between our prescription and the Casimir-like one. Let us notice that in the case where supersymmetry is linearly realized and so bosons and fermions are arranged in multiplets sharing the same
masses, this last term is zero, and so the two prescriptions give the same result. In the general case the difference manifests itself in the large $m$ limit.

By referring to the notations used in the paper the zero point energy reads $E_0(m)=(m_0/m)S(m)$ \footnote{Notice that our plane wave geometries are well defined only for $m_0\neq0$. This is evident for example from the definitions of the light-cone coordinates. The flat space case $m_0=0$ is so non continuously connected with our plane waves. In the flat space case the zero point energy is of course zero, a result independent on the way in which $E_0$ is evaluated. The parameter $m_0\neq 0$ can be reabsorbed by a redefinition of the light-cone coordinates and the only physical parameter to look at in our models is $m=m_0\alpha'p^+$. This explains why we study the behavior of $E_0$ as a function of $m$ and not of $m_0$.}. 
In the small $m$ limit only the zero-modes are leading order and $E_0(m)\to
m_0(\sum_i|b_i|-\sum_j|f_j|)$: this can be found by using both prescriptions. As a remark we noticed in the paper that for all the confining models we examined this quantity is negative. 

In the large $m$ limit, if we Casimir-renormalize each term in the sum by associating it with the term containing only the sum of Bessel functions, we get zero for the zero point energy. This is because the Bessel function vanishes exponentially for large values of the argument. 
Our prescription instead add to the Bessel terms the last line in (\ref{zerototus}), giving to $E_0$ a linear behavior in $m$, for large $m$. In the models we examined this reads as $E_0\approx -m_0A^2m$, where $A^2$ is a model-dependent constant.

As a remark, disconnected from the choice of the evaluation prescription for the zero point energy, and related to the models we examine in the main body of the paper, we should ask which kind of consequences the negativity of the
superstring zero point energy $E_0(m)$ should have at string level.

In the limit $m <<1$ (this is usually referred to as the
``supergravity limit'' as the leading order contribution to the string
motion comes from the zero-modes) we can study the consequences (or
the meaning) of the negativity of $E_0$ by referring to
supergravity. So, the question is: does the negative value of the
zero point energy correspond to some (classical) instability of the
backgrounds here considered?   In was shown in \cite{alfra2} that in
all the cases in which $E_0$ stays finite in the supergravity limit,
no instability can be read from the supergravity perspective, no
matter the sign of $E_0$. The situation is problematic only when $E_0$
goes to minus infinity for $m\to 0$ (just as in the Type 0B case
considered in \cite{alfra3}) where a tachyonic instability appears.

\section{Ricci flatness}\label{appendixd}
Imposing that the seven-dimensional metric
(\ref{ansatzCGLP}) is Ricci flat,
we find the following system of equations
\begin{align}
a'' = & \frac{1}{20 a^3 b^4 c^2 f^2}
[
-9 b^4 c^4 f^2 - 8 a^2 b^2 c^2 f^2 ( c^2 g^2 + 2 b^2( a'{}^2-1) )
\nonumber \\
- & 12 a^3 b^3 c f a' ( b f c' + c ( 2 f b' + b f'))
\nonumber \\
+ & 4 a^6 b^2 ( 3 f^2 g^2 + c^2 (3 g^2(1+g_3)^2 + 2 f^2 g'{}^2 ))
\nonumber \\
+ & a^4 c\, (c f^2 (f^2 + c^2(g^2+g_3)^2)
- 4 b^2 c f^2 (1 + g^2 - b'{}^2)
\nonumber \\
+ &
8 b^3 f b' ( f c' + c f' ) + b^4 (4 f c' f' - c^3 g_3'{}^2 ))
]
\\
b'' = & \frac{1}{20 a^4 b^3 c^2 f^2}
[ b^4 c^4 f^2 + 4 a^2 b^2 c^2 f^2 ( - 2 c^2 g^2 + b^2( a'{}^2-1 ))
\nonumber \\
+ & 8 a^3 b^3 c f a' ( b f c' + c ( -3 f b' + b f'))
\nonumber \\
- & 4 a^6 b^2 ( 2 f^2 g^2 + c^2 (2 g^2(1+g_3)^2 + 3 f^2 g'{}^2 ))
\nonumber \\
- & a^4 c\, ( 9 c f^2 (f^2 + c^2(g^2+g_3)^2)
- 16 b^2 c f^2 (1 + g^2 - b'{}^2)
\nonumber \\
+ &
12 b^3 f b' ( f c' + c f' ) + b^4 (-4 f c' f' + c^3 g_3'{}^2 ))
]
\\
c'' = & \frac{1}{20 a^4 b^4 c f^2}
[ 11 b^4 c^4 f^2 + 2 a^2 b^2 c^2 f^2 ( 11 c^2 g^2
+ 2 b^2( a'{}^2-1 ))
\nonumber \\
+ & 8 a^3 b^3 c f a' ( -4 b f c' + c ( 2 f b' + b f'))
\nonumber \\
+ & 2 a^6 b^2 ( -9 f^2 g^2 + c^2 ( g^2(1+g_3)^2 - f^2 g'{}^2 ))
\nonumber \\
+ & a^4 c\, ( c f^2 (f^2 + 11 c^2(g^2+g_3)^2)
- 4 b^2 c f^2 (1 + g^2 - b'{}^2)
\nonumber \\
+ &
8 b^3 f b' ( -4 f c' + c f' ) + b^4 (-16 f c' f' + 9 c^3 g_3'{}^2 ))
]
\\
f'' = & \frac{1}{20 a^4 b^4 c^2 f}
[ b^4 c^4 f^2 + 2 a^2 b^2 c^2 f^2 ( c^2 g^2 + 2 b^2( a'{}^2-1 ))
\nonumber \\
+ & 8 a^3 b^3 c f a' ( b f c' + 2 c ( f b' -2 b f'))
\nonumber \\
- & 2 a^6 b^2 ( - f^2 g^2 + c^2 (9 g^2(1+g_3)^2 + f^2 g'{}^2 ))
\nonumber \\
+ & a^4 c\, ( c f^2 ( 11 f^2 + c^2(g^2+g_3)^2)
- 4 b^2 c f^2 (1 + g^2 - b'{}^2)
\nonumber \\
+ &
8 b^3 f b' ( f c' - 4 c f' ) - b^4 (16 f c' f' + 11 c^3 g_3'{}^2 ))
]
\end{align}
\begin{align}
g'' = & \frac{c^2 g}{a^4} +
\frac{g (-2 b^2 + c^2 (g^2 + g_3))}{a^2 b^2}
+ \frac{f^2 g + c^2 g (1+g_3)^2 }{c^2 f^2}
\nonumber \\
- & \left( \frac{4 a'}{a}
+ \frac{c'}{c} + \frac{f'}{f} \right) g'
\end{align}
\begin{align}
g_3'' = & \frac{2 a^2 g^2 (1+g_3)}{b^2 c^2} +
\frac{f^2 (g^2 + g_3)}{b^4}
\nonumber \\
- & \left( \frac{2 a'}{a} + \frac{2 b'}{b}
+ \frac{3 c'}{c} - \frac{f'}{f} \right) g'_3
\end{align}
together with the constraint
\begin{align}
- \frac{1}{2 a^2 b^2 c f} &
[ b^4 c^4 f^2 + 2 a^2 b^2 c^2 f^2 ( c^2 g^2 + 2 b^2( a'{}^2-1 ))
\nonumber \\
+ & 8 a^3 b^3 c f a' ( b f c' + c ( 2 f b' + b f'))
\nonumber \\
+ & 2 a^6 b^2 ( f^2 g^2 + c^2 ( g^2 (1+g_3)^2 - f^2 g'{}^2 ))
\nonumber \\
+ & a^4 c\, (  c f^2 (f^2 + c^2(g^2+g_3)^2)
- 4 b^2 c f^2 (1 + g^2 - b'{}^2)
\nonumber \\
+ &
8 b^3 f b' ( f c' + c f' ) + b^4 ( 4 f c' f' - c^3 g_3'{}^2 ))
] = 0\,.
\end{align}
\noindent
The system is invariant under translations,
$r \to r + const$.

\noindent
Furthermore,
if $\{ a(r), b(r), c(r), f(r), g(r), g_3(r) \}$
is a solution of the equations of motion then
$\{ \lambda^{-1} a( \lambda r), \lambda^{-1} b(\lambda r),
\lambda^{-1} c( \lambda r), \lambda^{-1} f( \lambda r), g( \lambda r),
g_3( \lambda r) \}$ is a solution too $(\lambda \ne 0)$.
There is also a  symmetry under $g(r) \to -g(r)$.
Finally, in the case $g(r) \equiv 0$, there is
an extra  symmetry under $g_3(r) \to - g_3(r)$.


\newpage

\begin{figure}
\begin{center}
\epsfig{file=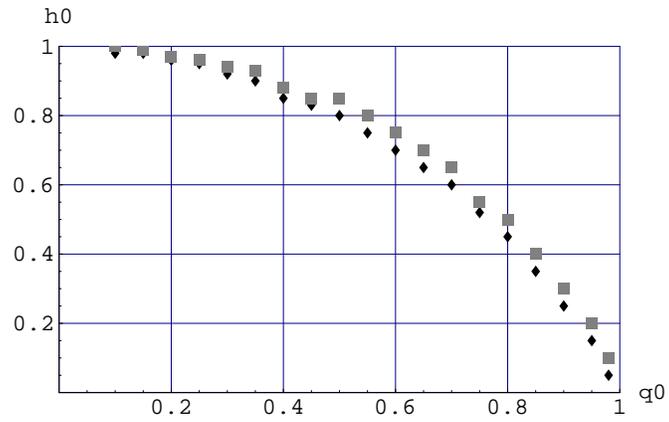,width=9cm}
\caption{ Every value of $h_0$ below a diamond
gives a regular solution, whereas
every value above a box leads to a singular one.
Also, recall that the full range is symmetric under $h_0 \to - h_0$.
\label{range}
}
\end{center}
\end{figure}

\begin{figure}
\begin{center}
\epsfig{file=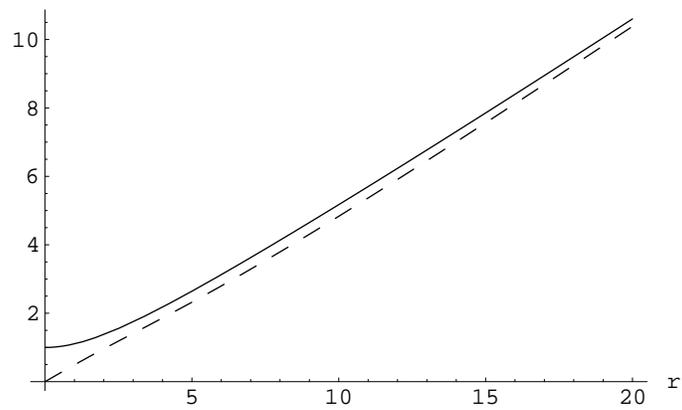,width=9cm}
\caption{ The functions $a$ (dashed line) and $b$ (continuous line).
\label{cippa1} }
\end{center}
\end{figure}

\begin{figure}
\begin{center}
\epsfig{file=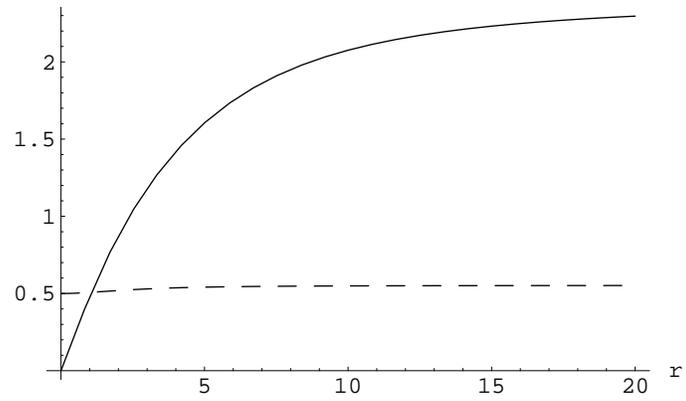,width=9cm}
\caption{ Plot of $-c$ (continuous line) and $f$ (dashed line).
\label{cippa3} }
\end{center}
\end{figure}

\begin{figure}
\begin{center}
\epsfig{file=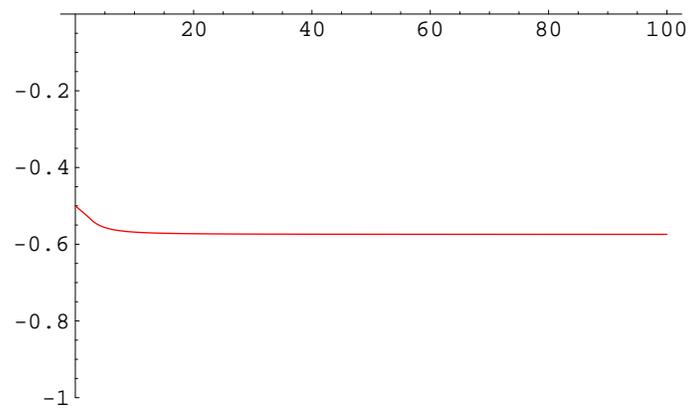,width=9cm}
\caption{ Plot of $g_3$.
\label{cippa4} }
\end{center}
\end{figure}

\begin{figure}
\begin{center}
\epsfig{file=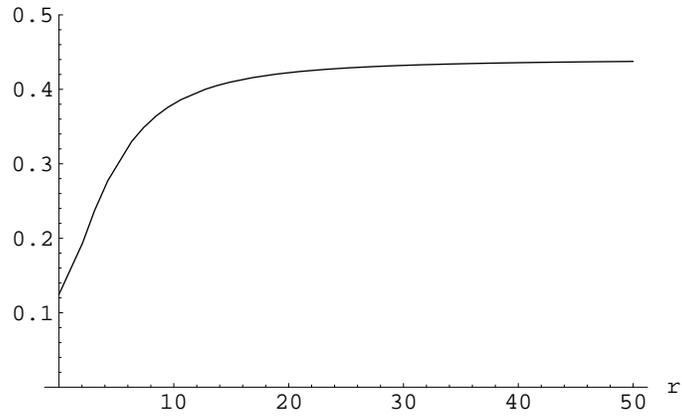,width=9cm}
\caption{ Plot of $e^\Phi$.
\label{cippa5} }
\end{center}
\end{figure}

\begin{figure}
\begin{center}
\epsfig{file=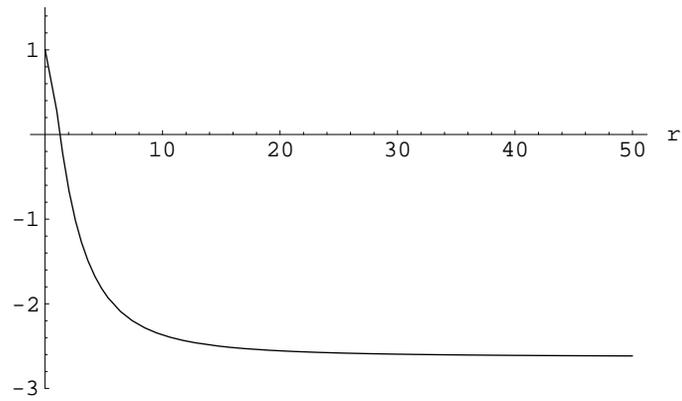,width=9cm}
\caption{ Plot of $A(r) \equiv \frac{f^2 - (1-g_3)^2 c^2}{f^2 + (1+g_3)^2 c^2}$.
\label{cippa6} }
\end{center}
\end{figure}

\begin{figure}
\begin{center}
\epsfig{file=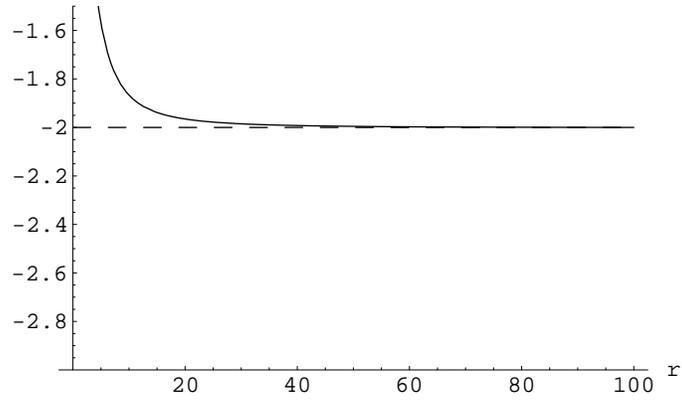,width=9cm}
\caption{ Plot of $A(r)$ for $\{ R_0=1, q_0=\frac{1}{2}, h_0=-0.395 \}$.
\label{cippa7} }
\end{center}
\end{figure}

\begin{figure}
\begin{center}
\epsfig{file=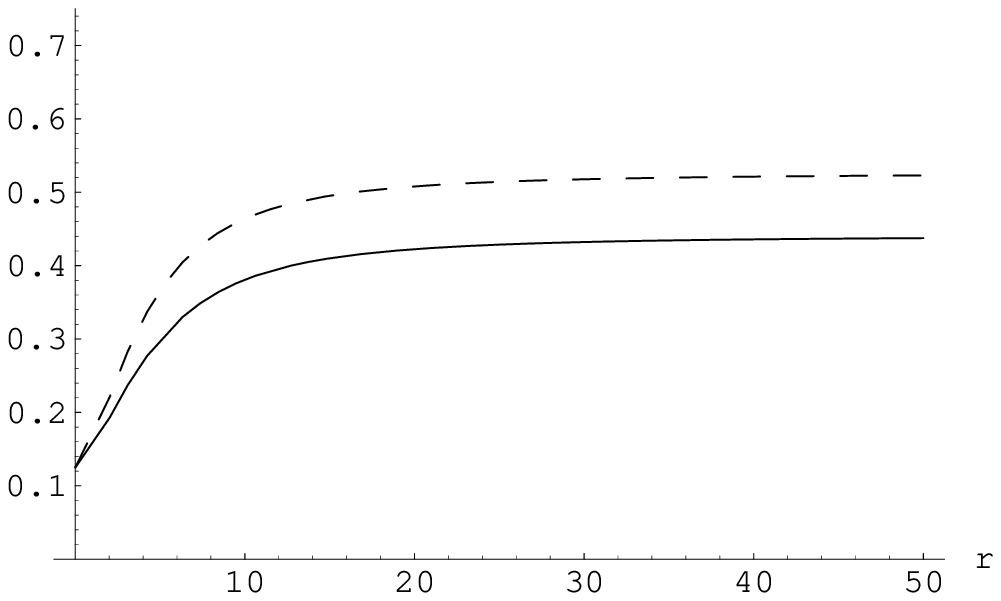,width=9cm}
\caption{ Plot of $e^\Phi$ for
$\{ R_0=1, q_0=\frac{1}{2}, h_0=-0.5 \}$
(continuous line) and
$\{ R_0=1, q_0=\frac{1}{2}, h_0=-0.395 \}$
(dashed line).
\label{cippa8} }
\end{center}
\end{figure}

\begin{figure}
\begin{center}
\epsfig{file=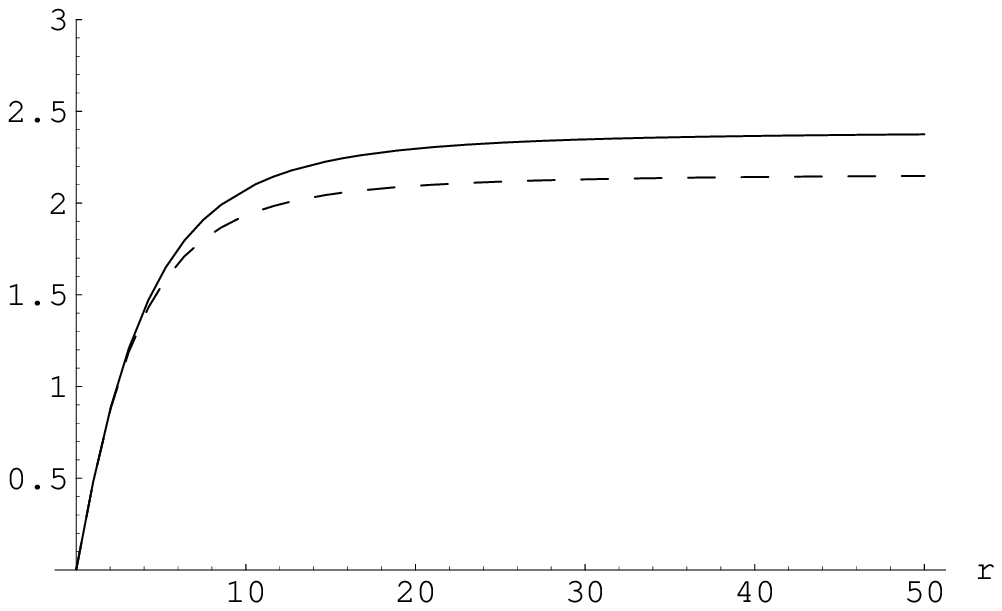,width=9cm}
\caption{ Plot of $-c$ for
$\{ R_0=1, q_0=\frac{1}{2}, h_0=-0.5 \}$
(continuous line) and
$\{ R_0=1, q_0=\frac{1}{2}, h_0=-0.395 \}$
(dashed line).
\label{cippa9} }
\end{center}
\end{figure}

\begin{figure}
\begin{center}
\epsfig{file=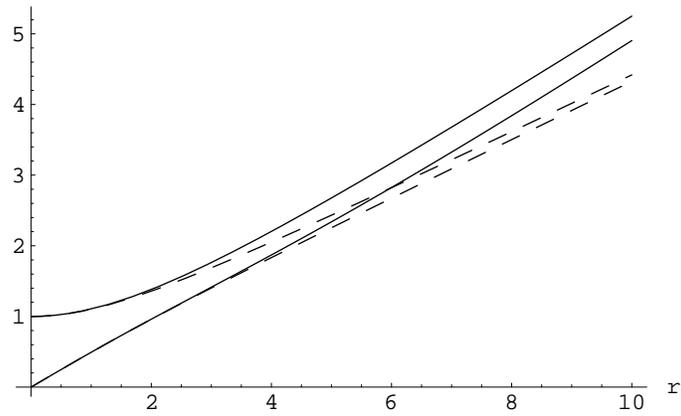,width=9cm}
\caption{ The functions $a$ and $b$ for the supersymmetric (dashed line)
and non-supersymmetric solution (continuous line)
$\{ R_0=1, q_0=\frac{1}{2} \}$.
\label{cippaAB} }
\end{center}
\end{figure}

\begin{figure}
\begin{center}
\epsfig{file=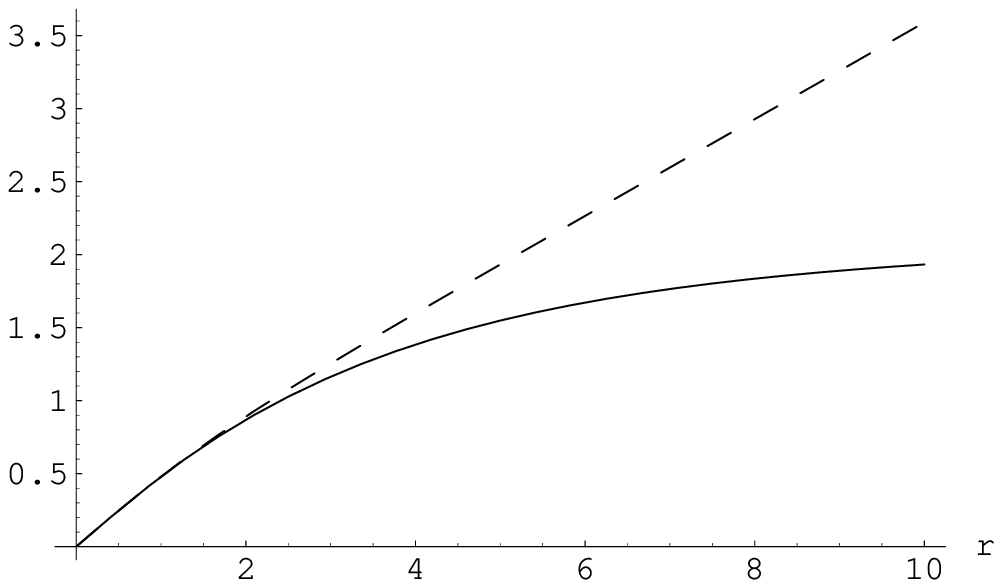,width=9cm}
\caption{ Plot of $-c$ for the supersymmetric (dashed line)
and non-supersymmetric solution (continuous line).
\label{cippac} }
\end{center}
\end{figure}

\begin{figure}
\begin{center}
\epsfig{file=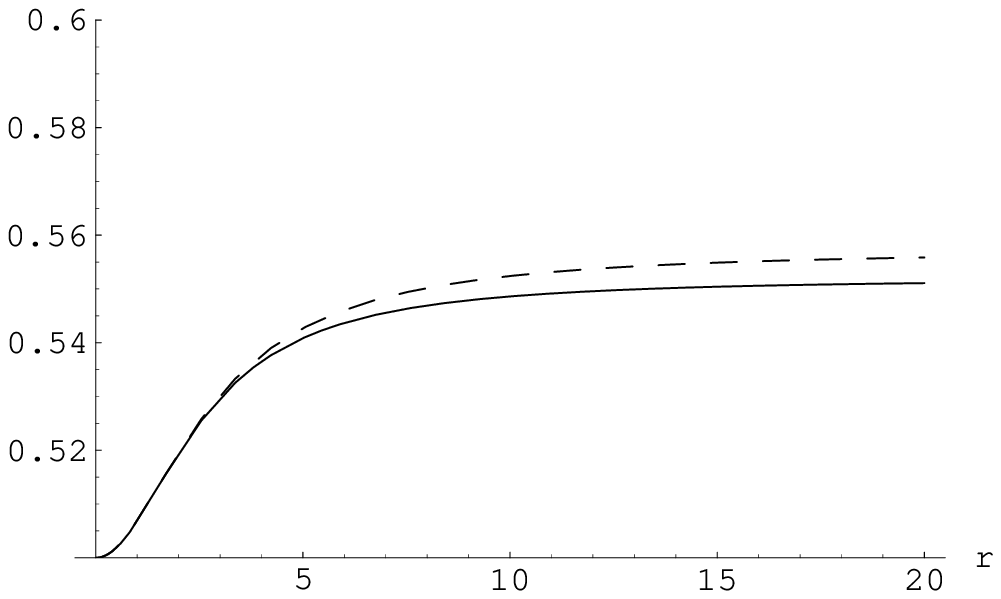,width=9cm}
\caption{ Plot of $f$ for the supersymmetric (dashed line)
and non-supersymmetric solution (continuous line).
\label{cippaf} }
\end{center}
\end{figure}

\begin{figure}
\begin{center}
\epsfig{file=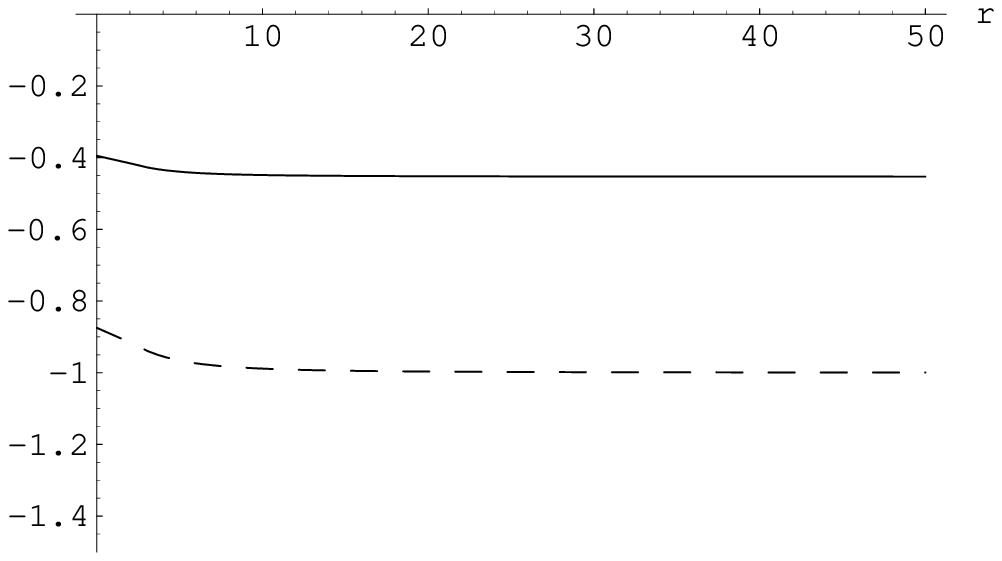,width=9cm}
\caption{ Plot of $g_3$ for the supersymmetric (dashed line)
and non-supersymmetric solution (continuous line).
\label{cippag3} }
\end{center}
\end{figure}

\begin{figure}
\begin{center}
\epsfig{file=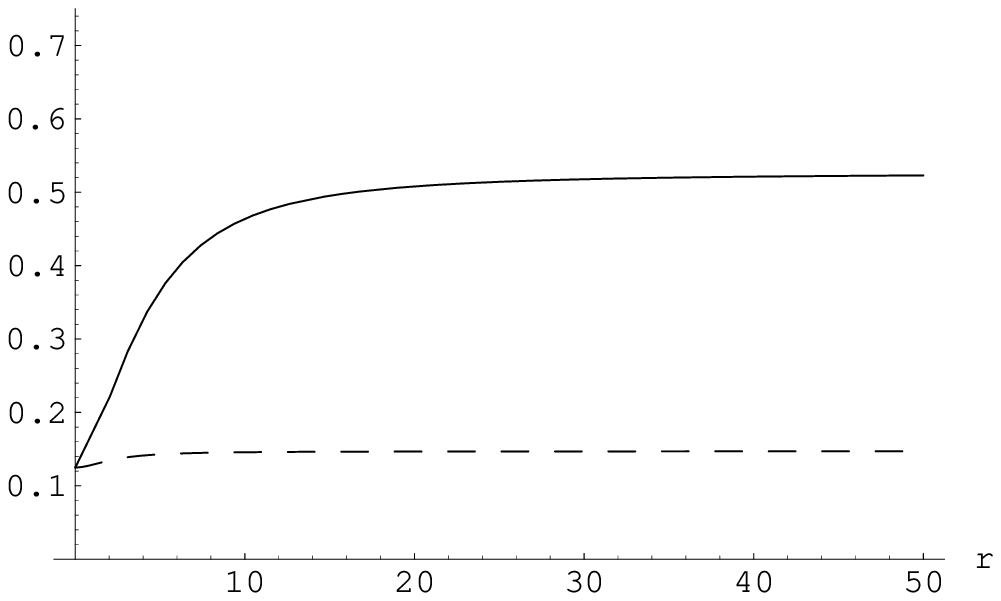,width=9cm}
\caption{ Plot of $e^\Phi$
for the supersymmetric (dashed line)
and non-supersymmetric solution (continuous line).
\label{cippaD} }
\end{center}
\end{figure}

\begin{figure}
\begin{center}
\epsfig{file=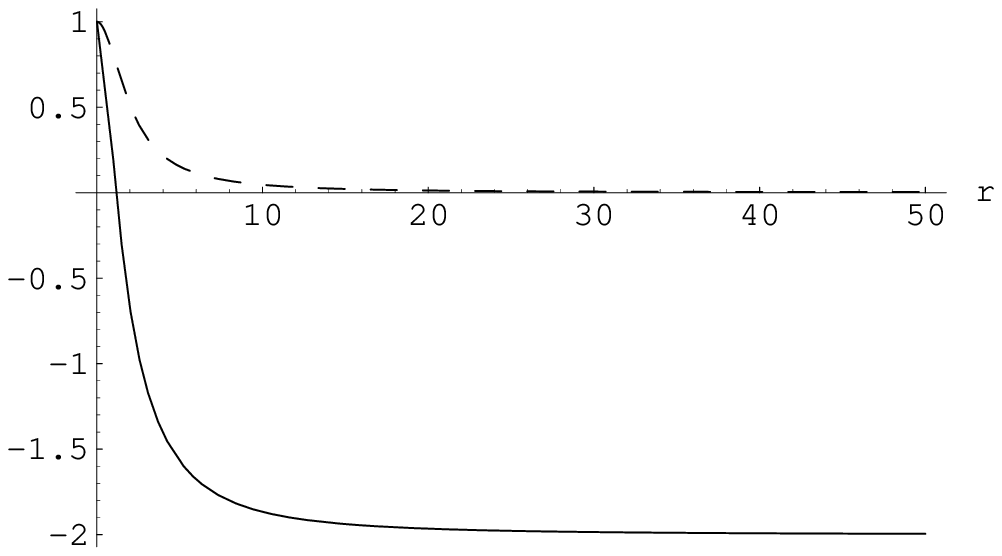,width=9cm}
\caption{ Plot of $A(r)$
for the supersymmetric (dashed line)
and non-supersymmetric solution (continuous line).
\label{cippaRR} }
\end{center}
\end{figure}

\end{document}